\documentclass[twocolumn,prd,nofootinbib,superscriptaddress,letterpaper]{revtex4}
\usepackage{hyperref}
\usepackage{amsmath}
\usepackage{graphicx}
\usepackage{color}
\usepackage{times}

\newcommand\mc{{{\cal M}_c}}
\newcommand\unit[1]{\, {\rm #1}}

\newcommand\qmstateproduct[2]{\left<#1|#2\right>}

\newcommand\ROS[1]{{\bf \color{blue} ROS: #1}}

\newcommand\hidetosubmit[1]{}
\newcommand\optional[1]{}
\newcommand\ForInternalReference[1]{}

\newcommand\abbrvACST{ACST}

\begin{document}
\title{Nonspinning searches for spinning BH-NS binaries in ground-based detector data: Amplitude and mismatch predictions in the constant
  precession cone approximation} 
\author{Duncan A. Brown}
\affiliation{Department of Physics, Syracuse University, Syracuse, NY 13244, USA.}
\author{Andrew Lundgren}
\affiliation{Department of Physics, Syracuse University, Syracuse, NY 13244, USA.}
\affiliation{Institute for Gravitation and the Cosmos, The Pennsylvania State University, University Park, PA 16802, USA.}
\affiliation{Albert-Einstein-Institut, Callinstr. 38, 30167 Hannover, Germany}
\author{R. O'Shaughnessy}
\affiliation{Institute for Gravitation and the Cosmos, The Pennsylvania State University, University Park, PA 16802, USA.}
\affiliation{Center for Gravitation and Cosmology, University of Wisconsin-Milwaukee, Milwaukee, WI 53211, USA}
\begin{abstract}
Current searches for compact binary mergers by ground-based gravitational-wave
detectors assume for simplicity the two bodies are not spinning.  If the
binary contains compact objects with significant spin, then this can reduce
the sensitivity of these searches, particularly for black hole--neutron
star binaries. In this paper we investigate the effect of neglecting
precession on the sensitivity of searches for spinning binaries using
non-spinning waveform models. We demonstrate that in the sensitive band of
Advanced LIGO, the angle between the binary's orbital angular momentum and its
total angular momentum is approximately constant. Under this \emph{constant
precession cone} approximation, we show that the gravitational-wave phasing is
modulated in two ways: a secular increase of the gravitational-wave phase due
to precession and an oscillation around this secular increase. We show that
this secular evolution occurs in precisely three ways, corresponding to
physically different apparent evolutions of the binary's precession about the
line of sight.  We estimate the best possible fitting factor between
\emph{any} non-precessing template model and a single precessing signal, in
the limit of a constant precession cone. Our closed form estimate of the
fitting-factor depends only the geometry of the in-band precession cone; it
does not depend explicitly on  binary parameters, detector response, or
details of either signal model.  The precessing black hole--neutron star waveforms least
accurately matched by nonspinning waveforms correspond to viewing geometries
where the precession cone sweeps the orbital plane repeatedly across the line
of sight, in an unfavorable polarization alignment.
\end{abstract}
\maketitle

\section{Introduction}
Ground based gravitational-wave detector networks (notably LIGO \cite{Abbott:2007kv,Shoemaker:aLIGO,2010CQGra..27h4006H} and Virgo
\cite{Accadia:2011zz,aVIRGO}) have performed several searches
for the inspiral and merger of binaries containing black holes (BH) and neutron stars (NS)
\cite{Abbott:2003pj,Abbott:2005pe,Abbott:2005qm,Abbott:2006bbh,Abbott:2007ai,Abbott:2007xi,Abbott:2009tt,Abbott:2009_12-18,Abadie:2011kd,Abadie:2011nz}.
The coalesence of NS-NS, BH-NS, and BH-BH binaries are the most
promising sources of gravitational-waves for these networks~\cite{Abadie:2010cf}.
For the lowest-mass compact binaries $M=m_1+m_2\lesssim 15 M_\odot$, the response of the detector to a binary merger  with arbitrary masses and
spins is well understood~\cite{%
gw-astro-mergers-approximations-SpinningPNHigherHarmonics,
2004PhRvD..70j4003B,BCV:PTF,2003PhRvD..67j4025B,2005PhRvD..72h4027B,%
2008PhRvD..78j4007H,%
Buonanno:2009zt%
}.
To identify the presence and properties of a signal,  the classic approach to data analysis on nonspinning binaries  has been matched filtering, linearly projecting the data
against each member of a
\emph{template bank}: a discrete array of signal models,  spaced to cover all possibilities with minimal inter-template
and template-signal
mismatch \cite{Owen:1995tm,Owen:1998dk,gw-astro-mergers-NonspinningTemplatesAreGoodForSpin-vdB2009,2007PhRvD..76j2004C,Allen:2005fk}. \optional{CITATIONS}
For \optional{incoherent searches for} nonspinning binaries, \optional{clarify for coincidence vs coherent?} the
templates needed lie in the two-dimensional signal manifold itself
(i.e., in $m_1,m_2$ or more naturally in the chirp mass $\mc=M \eta^{3/5}$ and symmetric mass ratio $\eta =m_1 m_2/M^2$)
\cite{Owen:1998dk,Owen:1995tm,2007PhRvD..76j2004C}.
For spinning binaries with more physical degrees of freedom, the added physical parameters
 require either (i) exact template banks that are impractically large, with a significant increase in computational
 burden and false alarm rate; or (ii) the use of an approximate \emph{detection template family}, whose waveforms
usually at best imperfectly approximate the target waveforms \optional{CITATIONS}
\cite{BuonannoChenVallisneri:2003a,2004PhRvD..70j4003B,BCV:PTF,2003PhRvD..67j4025B,2005PhRvD..72h4027B} but whose template banks
are smaller and better understood.
Though the more generic BCV detection template
family~\cite{2003PhRvD..67j4025B} has been applied in previous searches for
spinning systems\cite{Abbott:2007ai,2007CQGra..24.6227C}, recent studies
suggest that it is no more effective than the nonspinning template bank for
low-mass single-spin binaries
\cite{gw-astro-mergers-NonspinningTemplatesAreGoodForSpin-vdB2009}.   Current
templated searches for the gravitational-wave signature of low-mass merging
binaries are performed using this simple nonspinning
bank~\cite{Abbott:2009tt,Abbott:2009_12-18,Abadie:2011nz}.   
However, these results show that searching for generic spinning binaries with
a nonspinning search is sub-optimal and further work is needed to construct an
optimal search for spinning binaries in Advanced LIGO and Advanced Virgo. Such
a search may involve, e.g. a hierarchical combination of a nonspinning search
and a search using spinning templates, or a nonspinning search in some regions
of the parameter space, with targeted spinning searches in others. To devise
the correct stratergy, it is essential to understand where and why
advanced-detector searches will lose signal-to-noise ratio when searching for
spinning binaries using nonspinning waveforms; this is the goal of this paper.

Recent work on binaries where the spin is aligned with the orbital angular
momentum has demonstrated that advanced detectors are more sensitive to the
effects of spin on the orbital phase than first generation
detectors~\cite{2011PhRvD..84h4037A}. However, it has been shown that a
slightly more generic search would recover all nonprecessing signals well by
capturing the phase evolution of generic nonprecessing
sources~\cite{2011PhRvD..84h4037A}. 
Nonetheless,  a noticeable fraction of these  spinning binary mass, spin, and
orientation parameters $\lambda$ are at best poorly matched with nonspinning
and even generic nonprecessing templates, some with match lower than $\simeq
0.6$.  
These large mismatches  occur because nonspinning templates cannot capture the
effects of spin.   

This paper is concerned with binaries where the spin and angular momentum of
the binary are \emph{not} aligned. This induces \emph{precession}, causing the orbital angular momentum direction $\hat{L}$ changes with time. Because the instantaneous beam-pattern of the gravitational-wave emission is oriented along $\hat{L}$, the changing angle between $\hat{L}$ and the line of sight $\hat{n}$ produces large amplitude and phase modulations that are absent in
non-precessing signals.    Poor matches with nonspinning templates are disproportionately concentrated in systems that exhibit precession:
 systems with significant spin-orbit misalignment and asymmetric mass ratio. Conversely, binaries with tight spin-orbit alignment or simply limited variation in $\hat{L}\cdot\hat{n}$ are 
well-matched by non-precessing templates that capture  the effects of aligned spin \footnote{Aligned spin means that the spins of the compact objects are aligned with the orbital angular momentum. In this case there is no precession.} on the gravitational-wave phase
\cite{1993PhRvD..47.4183K,1995PhRvD..52..848P,2006PhRvD..74j4034B}.
Understanding the effect of precession is key to developing an optimal search
for spinning binaries in second-generation gravitational-wave detectors.

In this paper,  we investigate the  gravitational-wave signal from precessing  black hole--neutron star binaries.
Except for a short epoch of transitional precession (typically when $|\vec{L}|\simeq |\vec{S}|$; see \cite{ACST}), these binaries usually undergo \emph{simple precession}, where the orbital angular momentum
precesses around the total angular momentum with an ever-increasing opening angle \cite{ACST}.  
By investigating the spin evolution equations of \citet{ACST}, we find that  opening angle of the precession cone (i.e., the angle
between the total angular momentum $\vec{J}$ and the orbital angular momentum $\vec{L}$) does not expand significantly while the binary's signal is in the detector's sensitive band.
We perform analytic studies under the assumption that this opening angle is exactly constant; we call this the constant precession cone approximation.
We find the gravitational-wave phasing is modulated in two ways.
The first way that precession modifies the signal is via a secular change of the gravitational-wave phase.  In addition to the effect of orbital dynamics, the
gravitational-wave signal also accumulates a geometrical phase, a fraction of the ``precession phase'' of
$\vec{L}$ about $\vec{J}$.   
Previous work has demonstrated that precession induces a secular change in the phase accumulating over each precession cycle, 
 both in post-Newtonian studies \cite{ACST,BCV:PTF} and in numerical simulations of GR
 \cite{gwastro-mergers-nr-Alignment-BoyleHarald-2011}.  Though long  recognized, 
this factor is not included  in standard analytic non-precessing template models 
\cite{1993PhRvD..47.4183K,1995PhRvD..52..848P,2006PhRvD..74j4034B}.\footnote{The full effects of precession are of course captured in full time-domain calculations of orbit
  dynamics and precession.  These direct simulations, however, are presently too slow to be used in a search.
  Additionally, no one has constructed a complete template bank (a discrete set of reference signals) that adequately
  but minimally covers all possible signal options.}
We furthermore find this secular evolution occurs in three different ways, corresponding to physically different apparent evolutions
of the precession about the line of sight.
These three ways correspond precisely to whether the cone swept out by $\vec{L}$ encloses zero, one, or two
of the ``null lines'', which are the directions that produce exactly zero instantaneous response in the detector.   
The three regions correspond to whether  zero, one, or both directions are enclosed by the path swept out by $\vec{L}$ --
henceforth, the ``precession cone.''   
Each type  of configuration leads to a unique effect on the phasing of the binary.  If you are within one of the
regions, you accumulate secular phase at the same rate. 

The second way that precession modifies the signal is by introducing phase (and amplitude) modulations, on top of
secular evolution.   
These modulations cannot be captured by a non-precessing template.   A non-precessing signal model will therefore not
match a precessing signal optimally.   Moreover, the highest possible mismatch between a precessing signal and
non-precessing model can be estimated \emph{geometrically}.
In other words, our calculation tells us how easily a non-precessing search can find each particular precessing signal.
For simplicity, we perform our analytic studies assuming a source directly overhead a single detector.   
Our results will be generalized to a full network in a future analysis.   
Though simplified, our study is nonetheless
directly applicable to real searches.\footnote{For example, the two-detector LIGO network is nearly aligned and
  approximately sensitive to one polarization at a time.  In a subsequent
  paper we will address explicitly how to rescale our results to the two-detector LIGO network.}
 Moreover, given the complexity of realistic searches, special cases where searches have well-understood performance
provide invaluable tools for code validation and search calibration.  
Past, present, and future searches therefore benefit from the goal of this paper:  a simple model for how
well nonspinning searches recover most spinning, precessing low-mass binaries.

We will perform a detailed point-by-point numerical comparison of this expression to a large-scale Monte Carlo study of synthetic
searches in a subsequent
publication.   To compare with existing results in the literature, however, we apply our purely analytic results to
randomly selected binary parameters.   
We find our expressions   quantitatively agree with previous Monte Carlo studies of nonprecessing searches \cite{1995PhRvD..52..605A,2003PhRvD..67j4025B}, including  recent
investigations that include both mass and aligned spin as parameters \cite{2011PhRvD..84h4037A}.   For instance,
adopting the same  spin distribution as \cite{2011PhRvD..84h4037A}, we reproduce a similar distribution of fitting
factors. 
Unlike previous purely Monte Carlo studies, however, our analytic approach allows us to quantitatively predict precisely
which systems are not well fit with nonprecessing templates. The performance of a nonprecessing search on BH-NS systems with any mass and spin distribution can be easily predicted using our amplitude and mismatch predictions, and the simple predictions for the parameters that cause bad matches point the way toward improved searches.

In Section \ref{sec:DescribeTheory} we demonstrate that BH-NS binaries seen by gravitational-wave detectors will often
evolve on a constant precession cone.  %
In Section \ref{sec:CriticalOrientations} we 
parameterize the ways in which the binary's precession cone can be oriented relative to a single detector and its line
of sight to the source.    We find  that there are three distinct regions in this configuration space.  
In Section \ref{sec:Waveforms} we review the leading-order gravitational-wave emission from a precessing binary,
decomposing it into secular and modulated factors.   We show how the gravitational-wave signal from each region has a
distinctive secular dependence.   
Finally, in Section \ref{sec:Averaging} we average the amplitude over the precession cone, relating the mean power seen
along the line of sight to the power expected from an optimally oriented source.  Additionally, 
we argue that non-precessing signal templates necessarily cannot reproduce oscillations in the precessing signal model.
Using a similar average and assuming a generic signal model reproduces the \emph{secular} phase, we find an expression
for the  mismatch between non-precessing and precessing signals.

\section{Orbit dynamics and the constant precession cone limit}
\label{sec:DescribeTheory}

We consider a binary of two compact objects with masses $m_1$ and $m_2$, with $m_1 \geq m_2$, and intrinsic spins $\vec{S}_1$ and $\vec{S}_2$. From the coordinate separation $\vec{r}$ and the velocity $\vec{v} = \partial_t \vec{r}$, we can define the coordinate Newtonian angular momentum $\vec{L} = \mu \, \vec{r} \times \vec{v}$, where $\mu = \frac{m_1 m_2}{m_1 + m_2}$ is the reduced mass.

 For generic binaries, the orbit and spins evolve
according to complicated position-, spin-, and velocity-dependent expressions \cite{1993PhRvD..47.4183K,1995PhRvD..52..821K,2004PhRvD..70l4020S}.    Averaging these  post-Newtonian spin evolution
equations over a few orbital periods leads to  adiabatic expressions for the evolution of spin and orbital frequency
[see, e.g., \cite{ACST} (henceforth \abbrvACST{}), \citet{2004PhRvD..70l4020S}]: %
\begin{subequations}
\label{eq:SpinsEvolve}
\begin{eqnarray}
\partial_t \hat{L} &=& \vec{\Omega}_L \times \hat{L} \\
\partial_t \vec{S}_1 &=& \vec{\Omega}_{S_1} \times \vec{S}_1 \\
\partial_t \vec{S}_2 &=& \vec{\Omega}_{S_2} \times \vec{S}_2 \\
\vec{\Omega}_L &=& \frac{1}{r^3} \Bigg [  
  (2+ \frac{3 m_2}{2 m_1}) \vec{S}_1  + (2+ \frac{3 m_1}{2 m_2}) \vec{S}_2 \nonumber \\  
  && ~ - \frac{3}{2 |L|} [ (\vec{S}_2 \cdot \hat{L})\vec{S}_1 +  (\vec{S}_1 \cdot \hat{L})\vec{S}_2] \Bigg] \\
\vec{\Omega}_{S_1} &=& \frac{1}{r^3}\left[  
  (2+ \frac{3 m_2}{2 m_1}) \vec{L} + \frac{\vec{S}_{2}-3\hat{L} (\vec{S}_2\cdot \hat{L})}{2}
  \right] \\
  \vec{\Omega}_{S_2} &=& \frac{1}{r^3}\left[  
  (2+ \frac{3 m_1}{2 m_2}) \vec{L} + \frac{\vec{S}_{1}-3\hat{L} (\vec{S}_1\cdot \hat{L})}{2}
  \right]
\end{eqnarray}
\end{subequations}

In this expression we have omitted the quadrupole-monopole term described in \cite{2004PhRvD..70l4020S}, to more clearly
  correspond with the expressions from \abbrvACST{} on which we rely. We have included the spin-spin terms (containing both $\vec{S}_1$ and $\vec{S}_2$) for reference but we will soon assume that only one body is spinning.
These evolution equations cause $\vec{S}_1$ and $\vec{S}_2$ to precess but leave their magnitude constant. The direction $\hat{L}$ of the orbital angular momentum also precesses and gravitational-wave emission causes the length of $\vec{L}$ to shrink as the binary evolves.

 As described in \abbrvACST, when either one mass is nonspinning (e.g., $\vec{S}_2\simeq 0$),  the component masses
 are comparable ($m_1\simeq m_2$), or the smaller mass has a miniscule mass and necessarily spin ($m_1\gg m_2$), these expressions reduce to effectively
 \emph{single spin evolution}, meaning equations of motion are equivalent to the dynamics of $\vec{L}$ and a single (effective) spin.    For single spin binaries, spin-orbit evolution can occur in effectively two phases:
(i)  rare \emph{transitional precession} when $\vec{J}=\vec{L}+\vec{S}\simeq 0$, the relative change in $\vec{J}$ is large,
 and $\vec{J}$ can change direction; and  (ii)  \emph{simple precession}, when
are nearly constant $\vec{L}$ and $\vec{S}$ precess roughly steadily around $\vec{J}$.
Single-spin evolution is particularly appropriate for BH-NS binaries. Neutron stars likely do
not spin rapidly ($\vec{S}_2\approx 0$) -- observations suggest spin periods of no less than a few milliseconds
\cite{LorimerReviewCurrent}.
Observations and theoretical considerations constrain plausible NS masses to $1-3
M_\odot$,   ensuring the second object's minimal spin has
little impact on the orbit about an a priori higher-mass companion.
Albeit with less accuracy, BH-BH systems also often approximately satisfy the single-spin condition, as binaries that
are either comparable mass or dominated by a single spinning body occupy most of the
$m_1,m_2,\vec{S}_1,\vec{S}_2$ parameter space; see, e.g., the discussion in \cite{2004PhRvD..70j4003B}. 

\optional{
In particular, as

 - weak dependence on the orbital phase

 - typically only one spin (or effective spin) matters 

 - means that mass ($m_1,m_2$), spin magnitude $\chi_1$, three orientation angles $\theta_J,\beta^*,\psi_J$ (equivalent
 to $\kappa$) are needed: 6 parameters instead of the general 8  : THIS SHOULD BE BACK-PORTED
 
 - contrast with PTF: they attempt to directly maximize the 5-dimensional $2l+1$ spin polarization space
}

\ForInternalReference{
\ROS{Drafting notes}
* not always relevant (e.g., two spin cases) or strictly accurate (e.g., higher-order harmonics; etc); however, very helpful to interpret results.

* Recent citations:

Racine; Gergeley; etc

}

\subsection{Simple precession }
\begin{figure}
\includegraphics[width=\columnwidth]{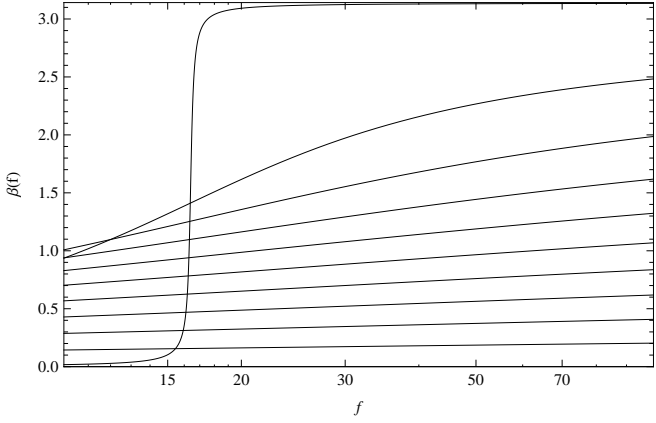}
\caption{\label{fig:Fiducial:BetaEvolve} \textbf{Constant precession cone approximation works}:  Opening angle
  $\beta(f)$   versus frequency for a $10 M_\odot+ 1.4 M_\odot$ binary with $\chi_1=1$ and
$\arccos \kappa =0,\pi/10, \ldots $ (bottom curves)  up to $\arccos\kappa \approx \pi$ (top, rapidly changing curve).  
We estimate $\beta(f)$ using Eqs. (\ref{eq:def:SingleSpin:beta:Evolve},\ref{eq:def:Gamma:PracticalEvolve}).
This plot shows that except for highly misaligned binaries
  ($\arccos\kappa$ large), $\beta(f)$
  is nearly constant, changing at most a fraction of a radian across the sensitive band of present and future
  detectors.  In general, only a small subset of   masses and highly misaligned spins are finely tuned enough to produce
  significant $\beta$ evolution in band.
}
\end{figure}

For the circular  orbits expected in astrophysical scenarios,   the  spin evolution  equations usually imply
\emph{simple precession}:   the orbital and spin angular momenta precess about the total angular
momentum, which  is fixed, except for negligible precession of its own about its average direction  [\abbrvACST{}].
Assuming $\vec{S}_2=0$, the system evolves with
$\kappa = \hat{L}\cdot \hat{S}_{1}$ and $ |S_{1}|  $ 
both  constant.  At early times, the total angular momentum  $\vec{J}$ is dominated by and closely aligned with
$\vec{L}$; as radiation carries away orbital angular momentum, eventually $\vec{J}$ is dominated by and closely aligned
with $\vec{S}$.  As the direction of the total angular momentum is nearly conserved, this process involves  substantial
change in the spin and orbital angular momentum directions.     Following \abbrvACST, we will employ the ratio of $\vec{S}_1$ to the Newtonian angular momentum
$\vec{L}=\mu \vec{r}\times \vec{v}$ as a parameter:
\begin{eqnarray}
\gamma(t) &\equiv& |\vec{S}_1|/|\vec{L}(t)| = \frac{\chi_1 m_1^2}{\eta M\sqrt{M r(t)} }
\end{eqnarray}
Using this parameter, the opening angle  angle $\beta$ of the precession cone (denoted $\lambda_L$ in \abbrvACST) can be expressed trigonometrically as 
\begin{eqnarray}
\label{eq:def:SingleSpin:beta:Evolve}
\beta(t)& \equiv& \arccos \hat{J}\cdot \hat{L} 
 =  \arccos \frac{1+\kappa \gamma}{\sqrt{1+2\kappa \gamma +\gamma^2}}
\end{eqnarray}
Note that for a spin-dominated binary ($\gamma \gg 1$)  $\kappa=\cos\beta$; for an angular momentum dominated binary
($\gamma \simeq 0$) $\beta\simeq 0$; and for a spin-orbit aligned binary ($\kappa=1$) all three of the spin, orbit, and
total angular momenta are aligned ($\beta=0$).  
At the level of accuracy of interest here,\footnote{As described at greater length in a subsequent publication, our
  constant precession cone approximation is designed to be accurate to of order several percent in power and mismatch.
  Quantitative comparisons of our approximation to numerical simulations will be provided in that publication.} the
opening angle   $\beta(f)$  at a given gravitational  frequency $f$ can be estimated by the above
expression, combined with the leading-order Newtonian expression  $r(f) = M(M\pi f)^{-2/3}$:
\begin{eqnarray}
\label{eq:def:Gamma:PracticalEvolve}
\gamma &\simeq& \chi_1 \frac{(m_1/M)^2}{\eta}(M\pi f)^{1/3} \nonumber \\
&\approx&  2.5 \chi_1 (m_1/10 m_2) \left(
   \frac{f M}{100 \unit{Hz} \;  10 M_\odot}
   \right)^{1/3}
\end{eqnarray}
Figure  \ref{fig:Fiducial:BetaEvolve} shows an example of $\beta(f)$ deduced from this expression.

During simple precession over longer timescales,  the width of the precession cone increases on the
gravitational-radiation timescale ($\tau_{gw}\simeq  [r^4/(\eta M^3)](5/64)  $).   During this slow increase, the orbital angular
momentum traces out an ever-widening spiral at  the  precession frequency: %
\begin{eqnarray}
\partial_t{\hat{L}}& =& \left(2+ \frac{3 m_2}{2 m_1}\right) \frac{\vec{J}}{r^3} \times \hat{L} \\
\vec{\Omega}_p &=& \left(2+ \frac{3 m_2}{2 m_1}\right) \frac{\vec{J}}{r^3} \\
|\Omega_p|&=& \left(2+ \frac{3 m_2}{2 m_1}\right)  \begin{cases}
\frac{\mu}{M^2} (M \pi f)^{5/3}   & |\vec{L}| \gg |\vec{S}| \\
\chi_1 \frac{m_1^2}{M^3} (M \pi f)^2      & |\vec{L}| \ll |\vec{S}|
\end{cases} 
\end{eqnarray}
where we have replaced $r$ by the leading-order PN expression $v^2= (M/r)=(M\pi f)^{-2/3}$ and $L_N=\mu
M(M\pi f)^{-1/3}$.  \abbrvACST{} provide an explicit, algebraic solution for the spins as a function of time [their Eqs. (59-63)].

\begin{figure}
\includegraphics[width=\columnwidth]{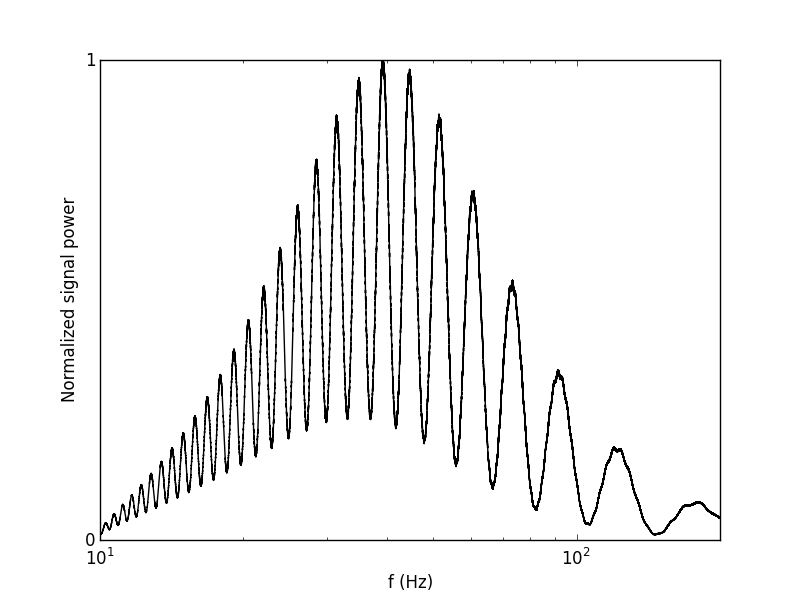}
\caption{\label{fig:Simple:AmplitudeModulationIsEasy}\textbf{Several precession cycles contribute comparably to the
    signal}: Power ($|\tilde{h}_\times|^2/S_h(f)$) versus frequency  for a spinning binary viewed edge
  on in the $\times$ polarization, computed directly from a Fourier transform of the gravitational-wave signal.    
In this figure, the binary is a $10+1.4 M_\odot$ BH-NS binary with $\chi_1=1$ and
  $\beta^*\approx \pi/3$ (solid). %
   Each pair of peaks corresponds to a single precession cycle. The signal is divided by the high-power zero-detune advanced LIGO noise PSD (power spectral density) \cite{LIGO-aLIGODesign-Sensitivity}.
}
\end{figure}

\subsection{Regions of parameter space I: L,S, dominated or intermediate }

\begin{figure}
\includegraphics[width=\columnwidth]{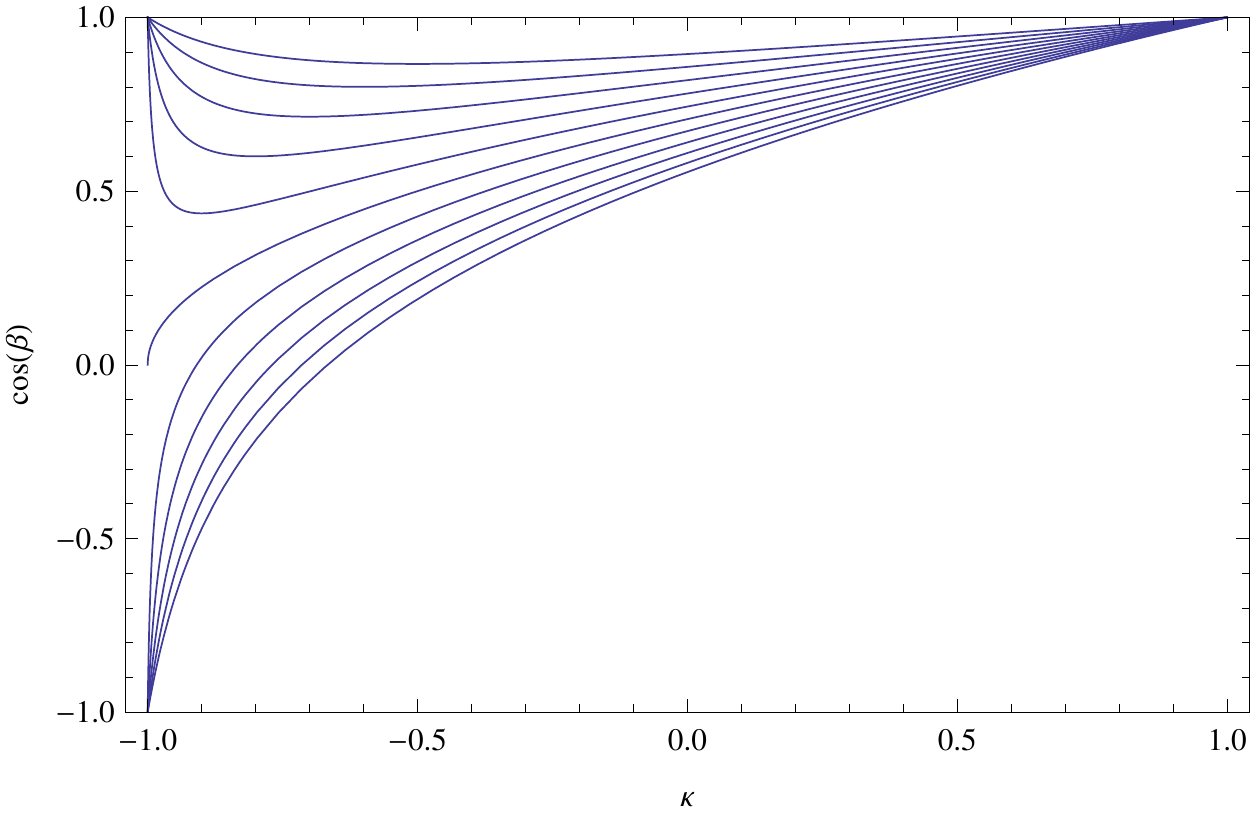}
\caption{\label{fig:SingleSpin:OpeningAngleRelation}\textbf{Precession opening angle versus spin-orbit misalignement}: Relation between  $\kappa=L.S$, and $\beta=\cos^{-1}L.J$ for
  $\gamma=|S|/|L| = 0.5,0.6\ldots 1.5$.  Binaries that are angular-momentum-dominated ($\gamma<1$) can have only a small range of precession cone opening
  angles, bounded by $\beta_{max}=\cos^{-1}\sqrt{1-\gamma^2}$.  Binaries that are spin-dominated ($\gamma>1$) can have
  all possible precession cone opening angles.  For extremely spin-dominated binaries $\kappa=\cos\beta$.
}
\end{figure}

\begin{figure}
\includegraphics[width=\columnwidth]{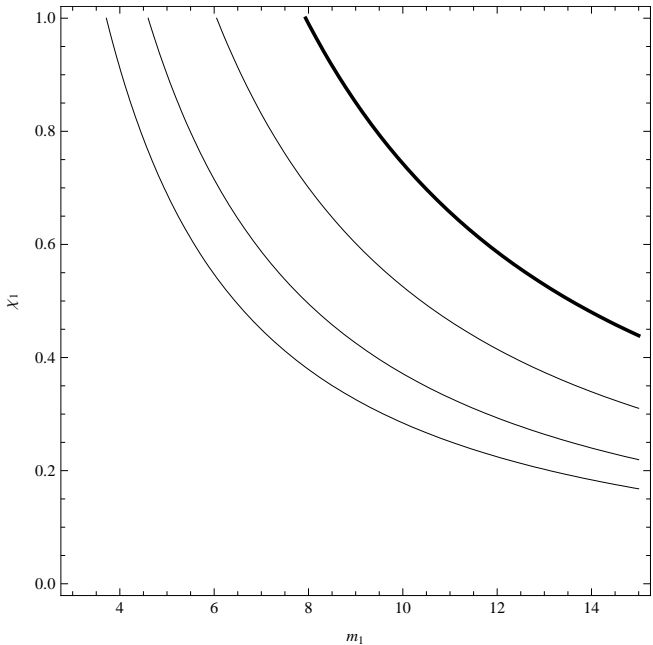}
\caption{\label{fig:BHNS:SimplePrecession:DominantRegions}\textbf{Angular momentum dominated versus spin-dominated binaries}:
In terms of the mass and spin of the black hole, contours of the ratio $\gamma=|\vec{S}|/|\vec{L}|$ evaluated at $40\unit{Hz}$.   The bottom left region is
angular-momentum dominated  ($|\vec{L}| \gg
|\vec{S}|$); the top right region of large black hole mass and spin is strongly spin-dominated ($|\vec{S}| \gg |\vec{L}|$).  Contours
show the ratio 
  $|\vec{S}|/|\vec{L}|\equiv\gamma=1$ (thick curve), $\sin \pi/4=1/\sqrt{2}$,  $\sin \pi/6 =1/2$ and
$\sin \pi/8$, evaluated  with a $1.4
M_\odot$ NS companion, versus the black hole mass and spin parameters
$m_{BH},\chi_{BH}$.  Above (below) the thick curve, BH-NS binaries' total angular momenta  are spin (orbit) dominated in
band.  If  spin and orbital angular momenta are nearly antialigned, these binaries have  undergone transitional precession at lower frequencies, typically
not in band.
Conversely, for orbital-angular-momentum-dominated binaries ($\gamma<1$), transitional precession has not occurred in
the past at lower frequencies and may, if anti-aligned and $\gamma$ near 1, occur in  band in the immediate future. 
Finally, below the bottom curve, BH-NS binaries waveforms are modulated little by precession in band. 
}
\end{figure}

Single-spin binaries can be loosely subdivided into rare transitional precession and ubiquitous simple precession [\abbrvACST].
In band, simple precessing binaries can be either $\vec{L}$ or $\vec{S}$ dominated.   A spin-dominated binary can have an arbitrary
opening angle.  An $\vec{L}$ dominated binary, by contrast, has a precession cone opening angle necessarily smaller than
$\pi/2$, and bounded above by 
\begin{eqnarray}
\beta_{max}\equiv  \sin^{-1} \gamma \; . %
\end{eqnarray}
In the neighborhood of this  extreme misalignment, at $\kappa=-\gamma$, the opening angle is nearly stationary with
spin-orbit misalignment  (i.e., $d\cos \beta/d\kappa\simeq0$).
In short, a distribution of $\vec{L}$ dominated binaries  has two choices for spin-orbit misaligment
(i.e., two values of $\kappa$) consistent with each realized opening angle.  Additionally, because of the local maximum
in $\beta$ as a function of $\kappa$, a randomly oriented distribution of spins will have opening angles $\beta$ that cluster
near that maximum  (i.e., $\beta\simeq \beta_{max}$).
To illustrate which regions are $\vec{L}$ and $\vec{S}$ dominated, Figure  \ref{fig:BHNS:SimplePrecession:DominantRegions} shows
contours of constant $\gamma$, assuming $m_2=1.4 M_\odot$.

\subsection{Regions of parameter space II: Steady precession and geometry}

\begin{figure}
\includegraphics[width=\columnwidth]{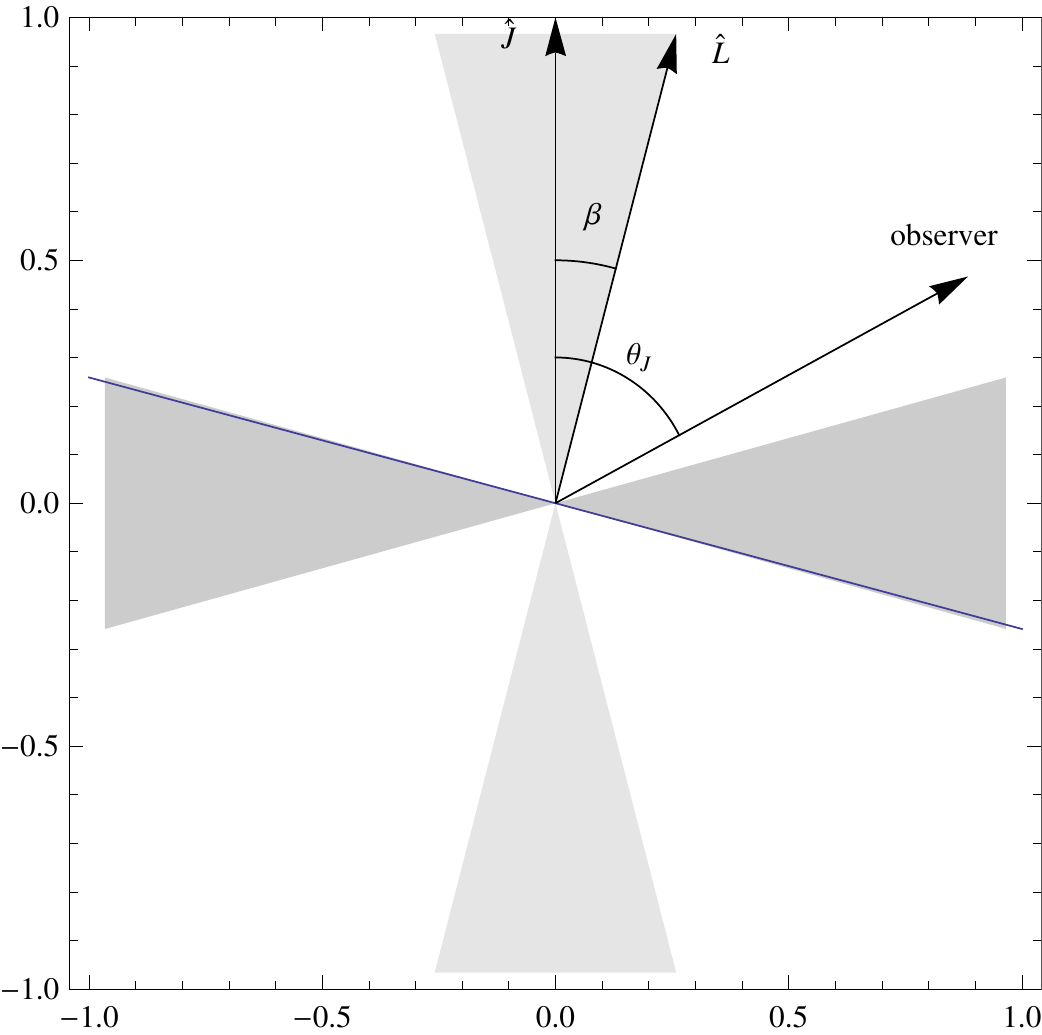}
\includegraphics[width=\columnwidth]{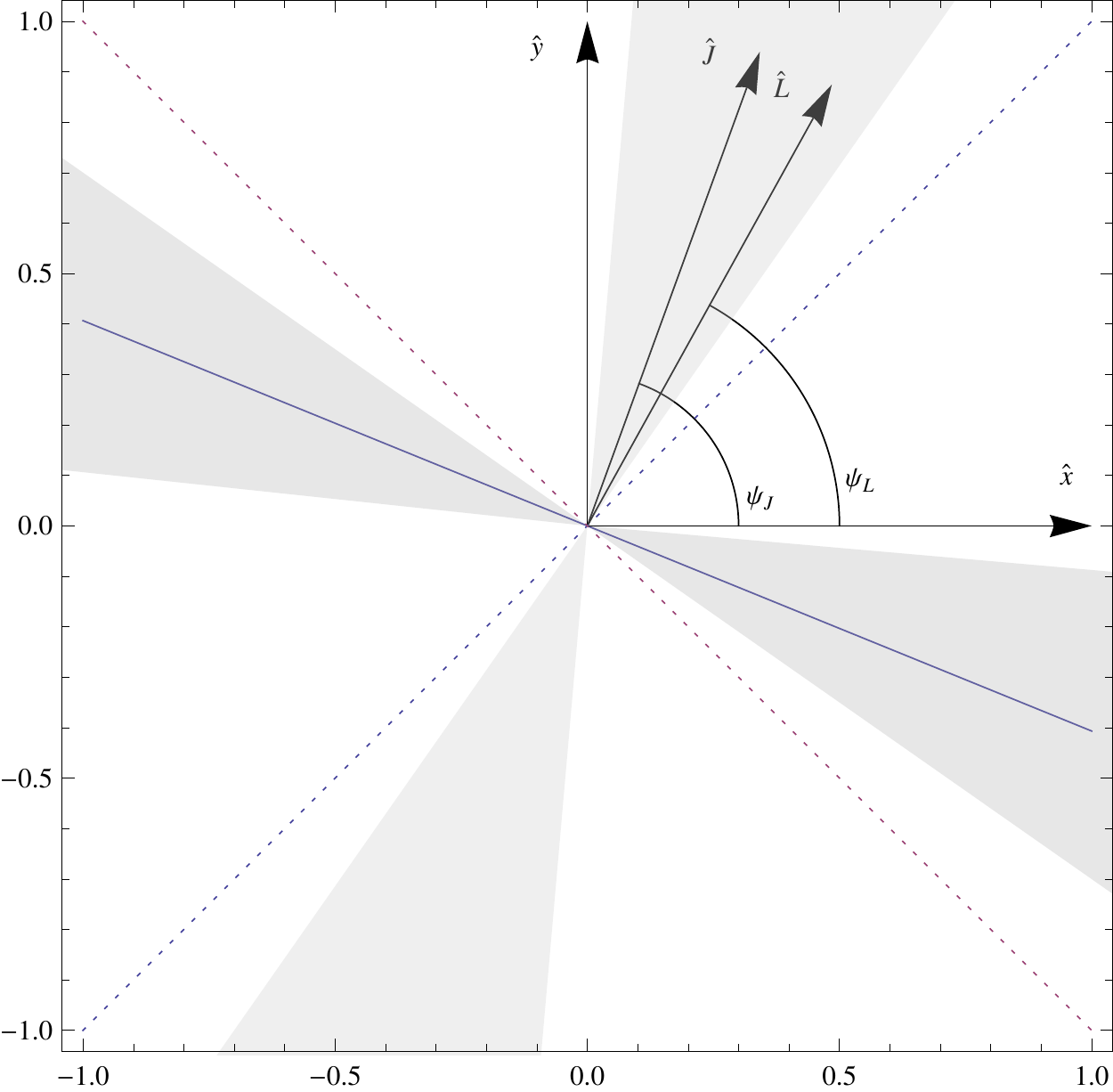}
\caption{\label{fig:BHNS:Geometry} \textbf{Coordinates for steady precession}: View from the side (top panel) and along
  the line of sight (bottom panel)  of the  geometry of a typical precessing BH-NS orbit, on timescales $t$ long compared with precession but short
  compared to gravitational-wave decay ($\Omega_p t\gg 1 \gg t/\tau_{gw}$).   The light shaded region indicates the cone
  swept out around the axis by the orbital angular momentum (the direction  of strong, circularly-polarized
  gravitational-wave emission from each orbit).  In both panels, the blue line indicates 
  a direction perpendicular to $\hat{L}$.  The dark shaded region indicates the corresponding region
  swept out by the orbital plane (the direction of weakest,  linearly-polarized emission); observers along these lines
  of sight see strong amplitude modulation.   The angle $\beta$ between the total angular momentum $\vec{J}$  and $\vec{L}$ is nearly constant on precession
  timescales.  On longer timescales, as $\beta$ increases, the line of sight to the observer is often enclosed in one of
  the two shaded regions.  On the bottom panel, the $\hat{x}$ and $\hat{y}$ axes correspond to the
  arms of the detector, projected into the plane of the sky.  The dotted lines correspond to ``null lines'', the
  directions that, when  $\hat{L}$ is parallel to them, produce zero amplitude in our detector; these lines lie in the
  plane of the sky, perpendicular to our line of sight.  %
[To distinguish $\vec{J}$ and $\vec{L}$ in the plane of the sky, the two panels adopt different reference times.]
}
\end{figure}

Unless transitional precession happens in band, ground-based
gravitational-wave detectors are sensitive to emission from a relatively well-defined epoch: the precession cone has
relatively constant opening angle [Fig. \ref{fig:Fiducial:BetaEvolve}].
    Quantitatively, we define a reference frequency $f_{peak}$
corresponding to the frequency  up to which half of the signal power has been accumulated. The specific reference frequency depends on the noise curve adopted.\footnote{In the text we choose the reference
  frequency as the half-power point, where $\int f^{-7/3}/S_h(f) df$ reaches half of its total value.
  Alternatively, the reference frequency can be set by maximizing $d\rho/d\ln f =  4
 f  |\tilde{h}(f)|^2/S_h$, or even  phenomenologically, in
  whatever manner is needed for numerically-calculated amplitude and match to reproduce our expressions.  For the noise
  curves considered in this paper, all   approaches nearly agree.}
For this paper, we adopt the fiducial advanced LIGO noise curve with zero-detuned signal recycling; see
\cite{LIGO-aLIGODesign-Sensitivity}). This includes a low-power mode for which $f_{peak} \simeq 40 \unit{Hz}$ and high-power for which $f_{peak} \simeq 60 \unit{Hz}$. However, all planned noise curves we have examined  have a reference frequency
in the neighborhood of which a constant precession cone is a good approximation.   
Henceforth the  ratio $\gamma=|S_1|/|L|$ and opening angle $\beta$ between $\hat{L}$ and $\hat{J}$  will refer to
quantities predicted  at this frequency by the simple precession expressions [Eqs. \ref{eq:def:Gamma:PracticalEvolve},\ref{eq:def:SingleSpin:beta:Evolve}].\footnote{For simplicity, we adopt the
  leading-order (Newtonian) expression for $r(f)$.  Higher order corrections are small. }

Second, not only is the precession cone nearly fixed, but as shown in Figure \ref{fig:Simple:AmplitudeModulationIsEasy} at least a few complete precession cycles occur 
between $20-100
\unit{Hz}$, where most of the signal-to-noise accumulates.  For example, for an angular-momentum-dominated binary ($\gamma \ll 1$), the number of
precession cycles for a single-spin binary can be approximated by the spin-independent expression
\begin{eqnarray}
N_P &\simeq&  \int_{\pi f_{min}}^{\pi f_{max}} d f_{orb} \frac{dt}{d f_{orb}} \Omega_p \nonumber \\
 &=&  \frac{5}{96}(2+1.5 \frac{m_2}{m_1}) [(M\pi f_{min})^{-1} - (M\pi f_{max})^{-1}] \nonumber \\
&\approx & \frac{27 (1+0.75 m_2/m_1) }{M/10 M_\odot} 
\end{eqnarray}
with a comparable but spin-dependent number for an S-dominated binary ($\gamma \gg 1$); see \abbrvACST{} Eqs. (45, 63) for a general solution.  As indicated by Figure \ref{fig:Simple:AmplitudeModulationIsEasy}, each precession cycle usually accumulates a comparable
proportion of detectable power (i.e., each pair of peaks is a similar order of magnitude in area).  More critically, the
figure indicates that at least one and often several precession cycles contribute to the total signal to noise. 
With many precession cycles, a gravitational-wave detector should be  relatively insensitive to the initial value of the
precession
phase.

For our purposes, then, the binary undergoes \emph{nearly steady simple precession} in band.  The instantaneous beam
pattern of the binary is aligned with the instantaneous $\hat{L}$, and is the same beam pattern as a non-spinning binary
\cite{1996PhRvD..54.4813W}. Along the axis aligned with $\pm \hat{L}$, the radiation is circularly polarized and in the orbital plane, the radiation is linearly polarized.  Over a longer timescale, $\hat{L}$ precesses around $\hat{J}$, sweeping the beam-pattern around the
precession cone.   Any given line of sight therefore can be characterized with its proximity to the orbital plane (an
amplitude minimum and specific, linear polarization) and $\pm\hat{L}$ (an amplitude maximum and specific circular
polarizations).   

\subsection{Constant precession cone coordinates}
\label{ap:SimplePrecessionCalculations}
In steady precession the unit vector $\hat{L}$ rotates
regularly about $\hat{J}$.   In practice, however, we interpret the gravitational-wave strain relative to a frame
associated with our line of sight $-\hat{n}$, where $\hat{n}$ is the emission direction from the binary.    To simplify
calculations, we adopt a frame defined by this direction: $\hat{z}=\hat{n}$, along with two perpendicular directions $\hat{x},\hat{y}$.
 Adopting an initial phase so the projection of $L$ into the plane of the sky (i.e.,
perpendicular to $n$) is along $\hat{J}$, the unit vectors needed to describe the steady precession approximation are
\begin{subequations}
\label{eq:Coordinates:J}
\begin{eqnarray}
\hat{J} &=& \sin \theta_J [\cos \psi_J \hat{x} + \sin \psi_J \hat{y}] + \cos\theta_J \hat{n} \\
\hat{p} &=& \frac{\hat{J}\times(\hat{J}\times n)}{|\hat{J}\times(\hat{J}\times n)|} \nonumber \\
&=& \cos \theta_J [ \cos \psi_J\hat{x} + \sin \psi_J \hat{y}] - \sin \theta_J \hat{n} 
\end{eqnarray}
Here $\psi_J$ is the orientation of $\hat{J}$ projected into the plane of the sky and $\theta_J$ is the angle between
$J$ and $\hat{n}$; see Figure \ref{fig:BHNS:Geometry}.
In terms of these vectors, the precession of the orbital angular momentum about $\hat{J}$ can be described as
\begin{eqnarray}
\hat{L} &=& \cos \beta \hat{J} +  \sin \beta R_{\hat{J}}(\alpha) \hat{p}  
\end{eqnarray}
where $R_{\hat{J}}(\alpha)$ is a rotation operator about $\hat{J}$ and   $\alpha=\int
\Omega_p dt$ is the precession phase of $\vec{L}$ around that axis \cite{ACST}.

The gravitational-wave strain recovered by a detector depends on the orientation of $\hat{L}$ relative to the radiation
frame.  In terms of the coordinates above, the inner products needed are
\begin{eqnarray}
\hat{L}\cdot \hat{n} &=& \cos \beta \cos \theta_J - \sin\beta \sin \theta_J \cos \alpha \\
\hat{L}\cdot\hat{x} +i \hat{L}\cdot \hat{y}&=& e^{i\psi_J}[ \sin\beta(\cos\alpha \cos \theta_J + i \sin \alpha) 
  \nonumber \\
 &+ &\cos
  \beta \sin \theta_J] \\
 &=& e^{i \psi_L} \sqrt{1-(\hat{L}\cdot\hat{n})^2}
\end{eqnarray}
\end{subequations}
where $\psi_L$ is shorthand for the instantaneous orientation of $\hat{L}$ on the plane of the sky.   

For convenience and without loss of generality, in what follows we will
assume  the binary is directly overhead a single interferometer, with arms along $\hat{x}$ and $\hat{y}$.  
For this orientation,  the detector response functions are $F_+=\cos 2\psi_J,F_\times=\sin 2\psi_J$.  
  If the source is not directly overhead, the single-detector response functions
\begin{subequations}
\begin{align}
F_+ &= \frac{1}{2} (1 + \cos^2 \theta) \cos 2\phi \cos 2\psi - \cos \theta \sin 2\phi \sin 2\psi \\
F_\times &= \cos \theta \sin 2\phi \cos 2\psi + \frac{1}{2} (1 + \cos^2 \theta) \cos 2\phi \sin 2\psi
\end{align}
can be rewritten as an overall scaling and a polarization shift
\begin{align}
F_+ &= F_0 \cos 2 (\psi + \psi_0) \\
F_\times &= F_0 \sin 2 (\psi + \psi_0)
\end{align}
with
\begin{gather}
F_0 = \sqrt{((1+\cos^2 \theta)/2)^2 \cos^2 2\phi + \cos^2 \theta \sin^2 2\phi} \\
\tan 2\psi_0 = \frac{\cos \theta}{(1+\cos^2 \theta)/2} \tan 2\phi ~.
\end{gather}
\end{subequations}

\section{Critical viewing orientations and domains}
\label{sec:CriticalOrientations}
The instantaneous emission pattern of gravitational radiation is aligned with the instantaneous orbital angular momentum $L$ and is the same as the beam pattern of a non-spinning binary i.e. an $l=|m|=2$ quadrupolar configuration. The radiation is circularly polarized along $L$ and $-L$, and linearly polarized in the plane perpendicular to $L$.    In particular,
when the orbital angular momentum is instantaneously perpendicular
to the line of sight, one (linear) polarization of the gravitational-wave signal is instantaneously zero.  
The condition that one polarization be zero at some instant defines a two-dimensional surface in the three-dimensional
space  ($\theta_J,\psi_J,\beta$)  of all possible precession cone geometries.  As we show below, this surface decomposes
the three-dimensional space into three distinct regions, corresponding to different ways the orbital plane crosses the
line of sight or, equivalently, different ways the orbital angular momentum wraps around two null lines.

\begin{figure*}
\includegraphics[width=0.3\textwidth]{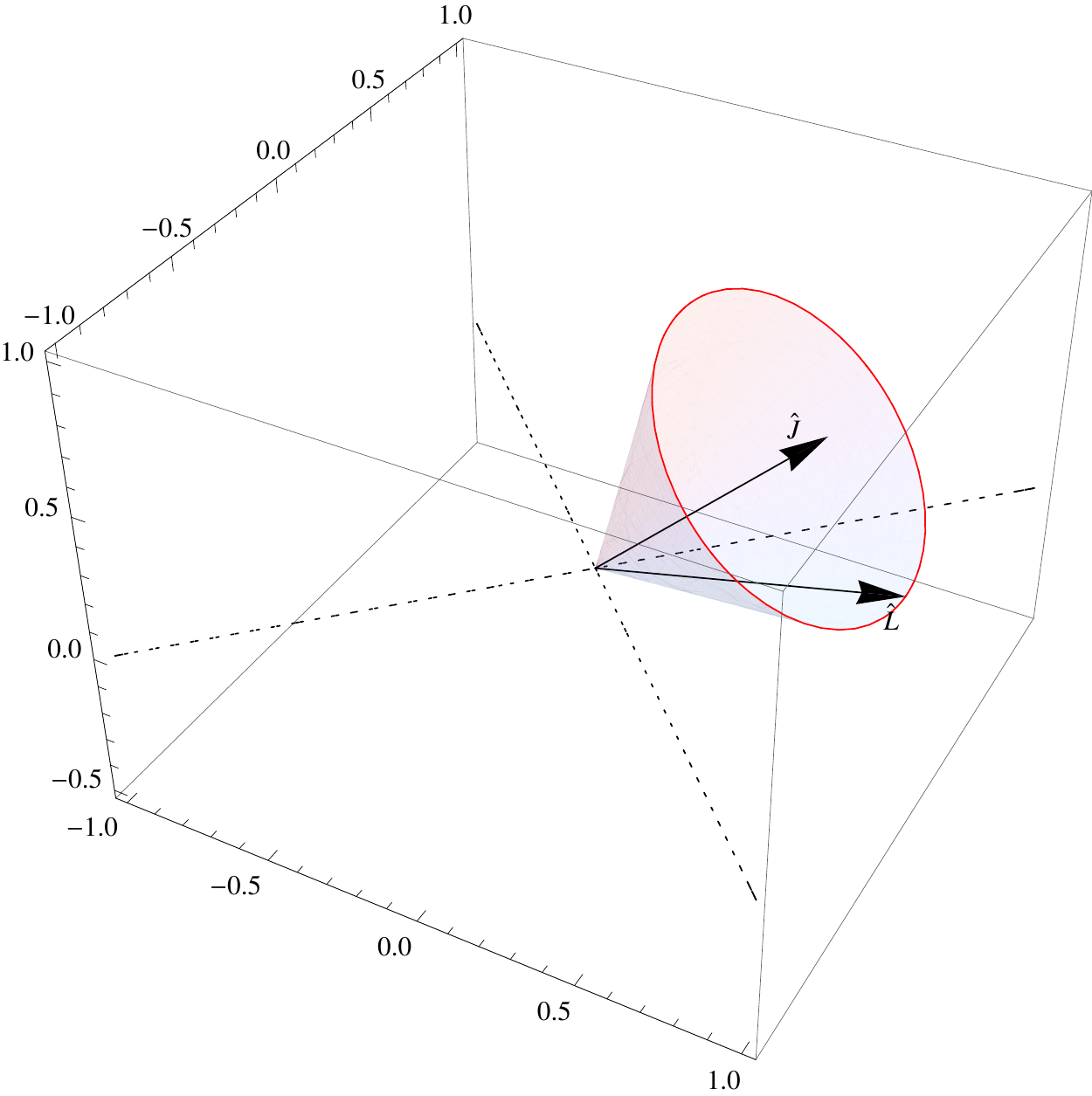}
\includegraphics[width=0.3\textwidth]{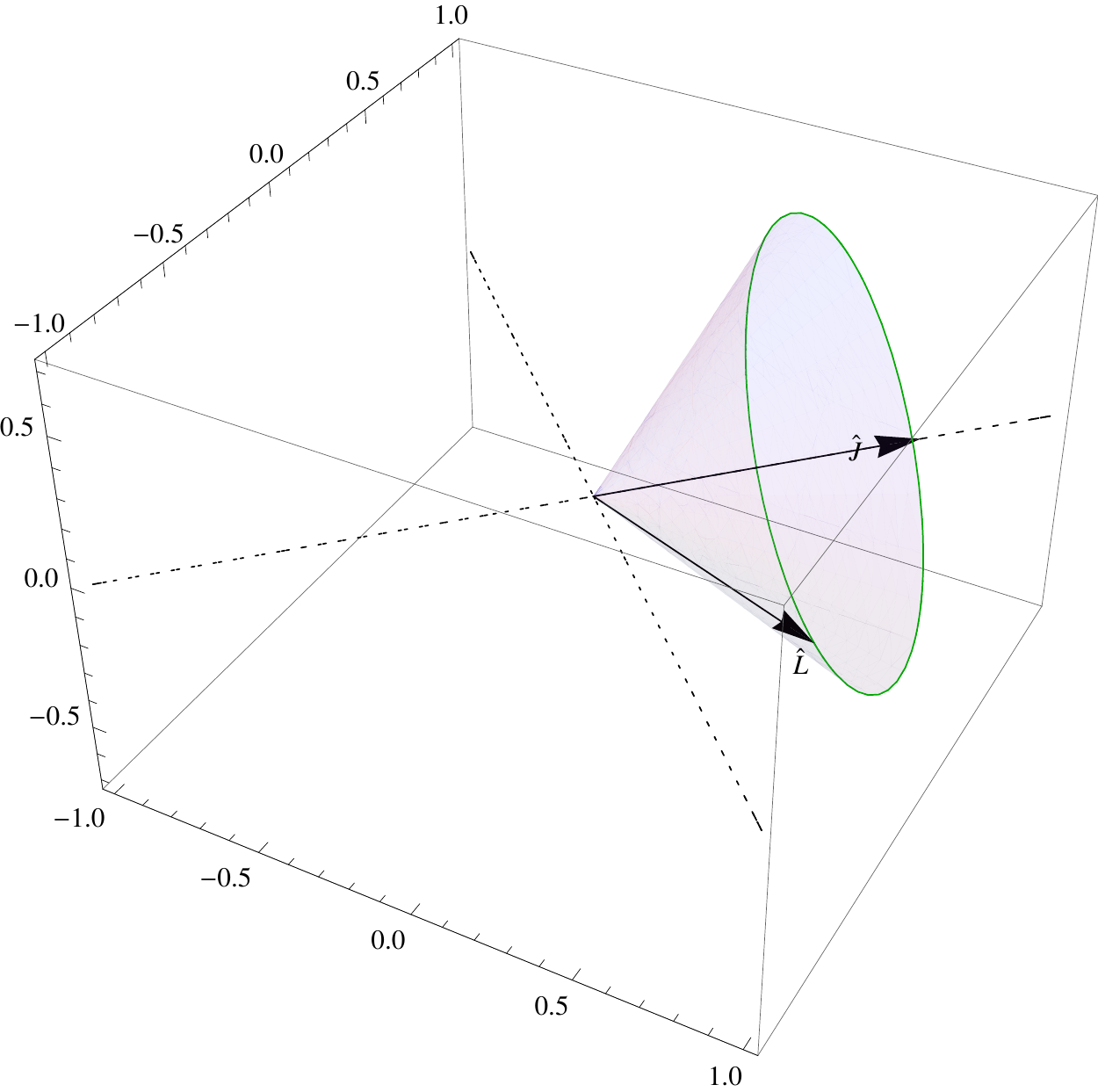}
\includegraphics[width=0.3\textwidth]{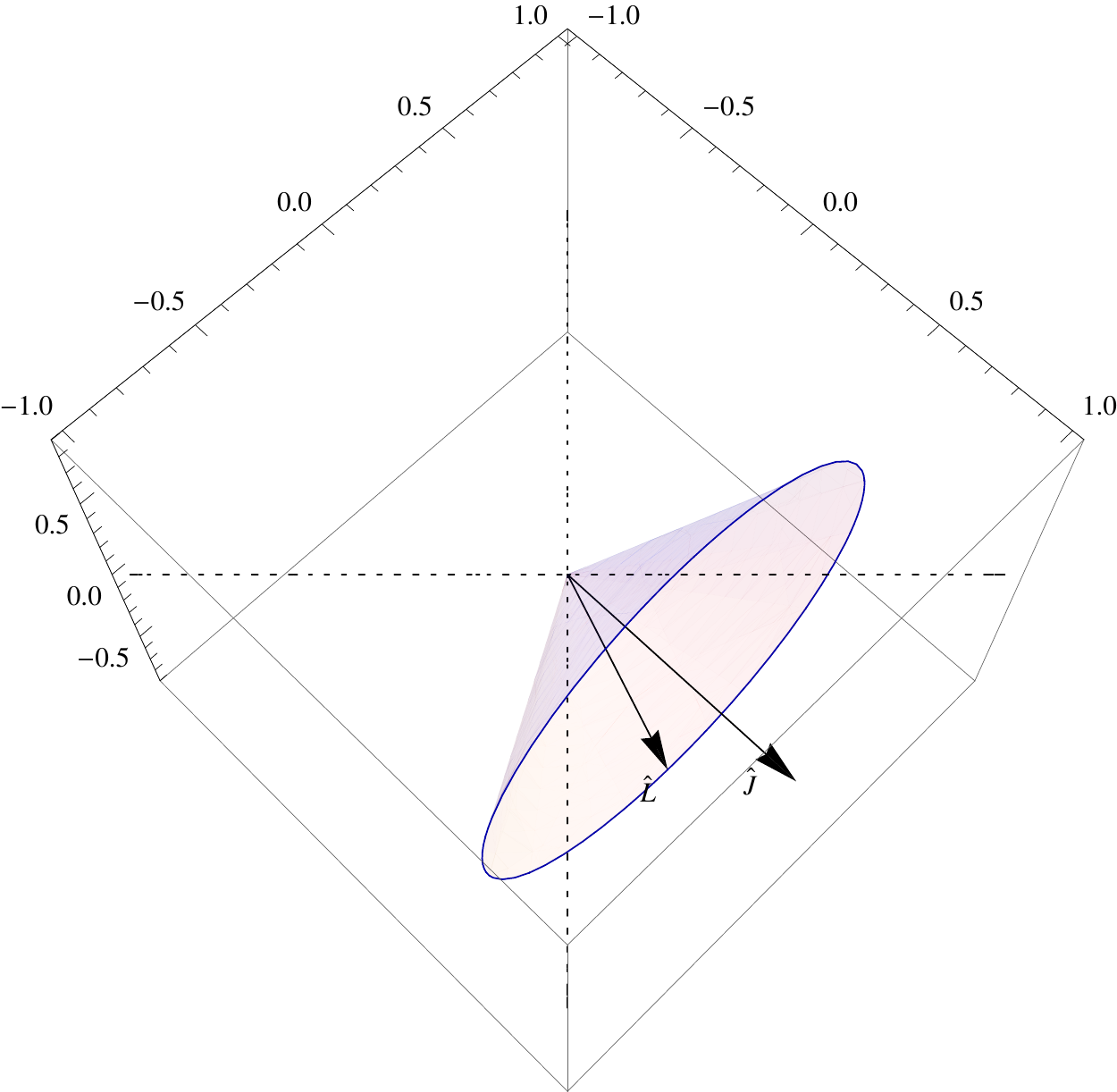}
\caption{\label{fig:ConeDemo}\textbf{Three regions have three types of precession cones}: The figure above shows the
  precession cone for three fiducial binaries, characterized by the angles $(\theta_J,\beta,\psi_J)=(\pi/4,\pi/6,0), (\pi/2, \pi/4,\pi/4)$ and
  $(\pi/2,\pi/3,0)$.  Each figure shows the precession-induced path of $\hat{L}(t)$ (colored circles), the total angular momentum direction (arrow), and
  an example of an instantaneous $L$ direction (arrow).  As these three examples demonstrate, each region corresponds to a different way the precession
  cone wraps around the two dotted lines. These two lines correspond to orientations
  of $\hat{L}$ such that the detector would instantaneously see exactly zero amplitude. We will henceforth denote them as ``null lines''. 
These three precession cones are representative examples of the three regions discussed in the text:  R1 (left panel), R2 (center panel) and R3 (right
  panel)
}
\end{figure*}

The detector is not sensitive to (leading-order) gravitational radiation emitted from the binary when the orbital
angular momentum simultaneously satisfies the following two conditions: 
\begin{subequations}
\label{eq:TransverseCondition}
\begin{eqnarray}
0&=&\hat{L}\cdot \hat{n} =0 \\
0&=&(\hat{L}\cdot \hat{x})^2 - (\hat{L}\cdot\hat{y})^2  
\end{eqnarray}
\end{subequations}
Solutions to this expressions exist if and only if the angle between $\vec{J}$ and one of the four critical directions indicated in Figure \ref{fig:ConeDemo} is equal to the angle $\beta$ between $\vec{J}$ and $\vec{L}$, or equivalently if the following expression is zero:
\begin{eqnarray}
d_{null} &\equiv& (\hat{J} \cdot v_1- \cos \beta)(\hat{J}\cdot v_1 + \cos\beta) \nonumber \\
 &\times &(\hat{J}\cdot v_2 -\cos \beta)(\hat{J}\cdot v_2 + \cos \beta) \\
  &=&  [(\hat{J}\cdot v_1)^2 - \cos^2 \beta] [(\hat{J}\cdot v_2)^2 - \cos^2 \beta] 
\end{eqnarray}
where $v_1 = (\hat{x}+\hat{y})/\sqrt{2}$ and $v_2 = (\hat{x}-\hat{y})/\sqrt{2}$.    Substituting in the coordinate form
for $\hat{J}$ given above and solving $d_{null}=0$ for $\cos^2 2\psi$ leads to an expression for the boundary
polarization angle $\psi_{J,crit}$ between different regions:
\begin{eqnarray}
\label{eq:def:dnull:zeros}
\cos^2(2\psi_{J,crit}) &=& \frac{4 \cos^2 \beta(\cos^2\beta-1)}{\sin^4 \theta}  \nonumber \\
 &=&  - \frac{(1+\cos2\beta)(\cos 2\beta + \cos 2\theta)}{\sin^4 \theta}
\end{eqnarray}

\begin{figure}
\includegraphics[width=\columnwidth]{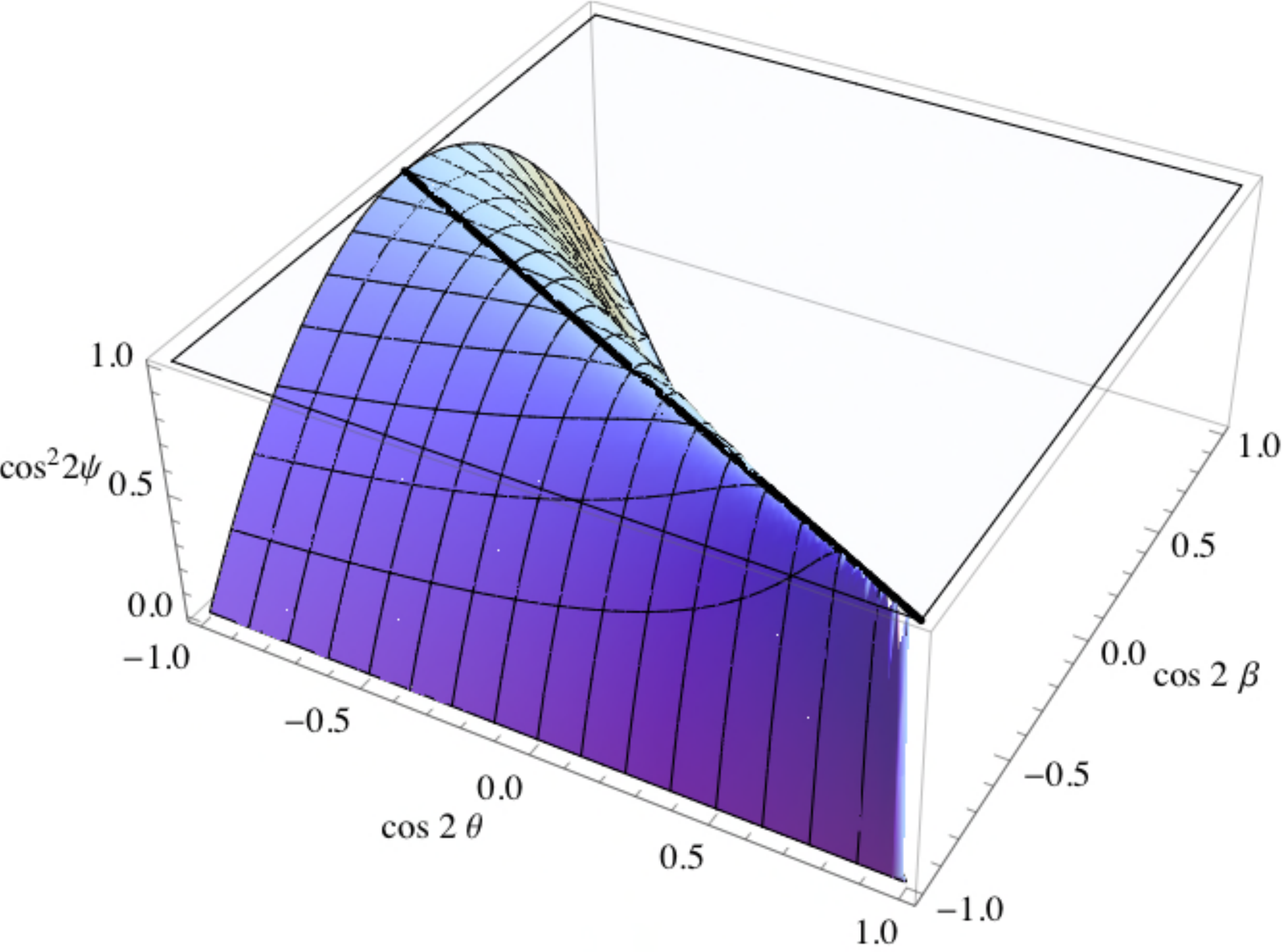}
\caption{\label{fig:RegionDividers}\textbf{Region divisions in coordinate form}:  The three ways the binary's orbital
  angular momentum can wind around the two null lines are separated by cases where the cone is tangent to one or the
  other axis.  In this figure we express this dividing surface using explicit coordinates for the 3d configuration
  space: $z=\cos^2 2\psi_J$,  $x=\cos
  2\theta_J$, $y=\cos 2 \beta$.   Points on this surface are simultaneous solutions to 
  Eq. (\ref{eq:TransverseCondition}), as calculated by Eq. (\ref{eq:def:dnull:zeros}).   In this figure, the R1 region
  is the largest, farthest from the viewer; the region R3 is the region above the surface, closest to the viewer; and the region R2 is below
  the surface.
}
\end{figure}

\begin{figure}
\includegraphics[width=\columnwidth]{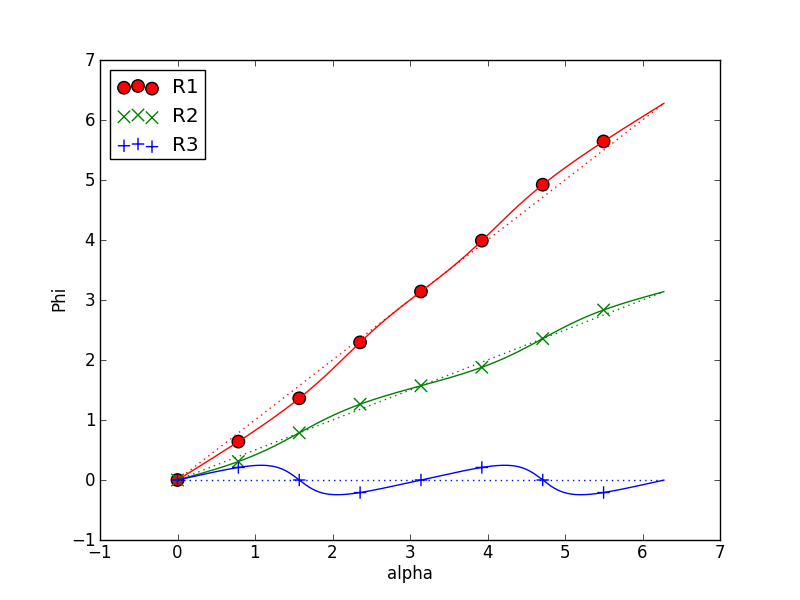}
\caption{\label{fig:ThreeCanonicalPhasings} \textbf{Phase evolution versus time}: For three selected viewing geometries
  of a single binary with $\beta=60^o$, a
  plot of $\Phi-\Phi_{S}(t) = \text{arg}(z)$ versus $\alpha$, the precession phase.    This plot shows how much of the precession phase
  is accumulates in the gravitational-wave signal due to precession  
  [Eq. (\ref{eq:PhaseDecomposition})].   These three phase evolutions are generated from Eqs. (\ref{eq:h:ViaCpSpEtc}).
  The dotted line corresponds to $W\alpha$, where $W$ corresponds to the rate of secular phase increase appropriate to
  that line of sight.
The three viewing geometries lie in the three   regions R1,R2,R3 described in the text.
}
\end{figure}

These configurations defined above divide the space of orientations $(\theta_J,\beta,\psi_J)$ into five regions, henceforth
denoted R1$_\pm$, R2$_\pm$, and R3 [Figure \ref{fig:RegionDividers}].   As our notation suggests, using symmetry we can
relate two pairs of regions by reflection symmetry $\beta \rightarrow \beta-\pi$.   The region R1 (the union of R1$_{+}$
and R1$_{-}$)  
corresponds to all configurations such that the orbital angular momentum does not wind around either null line.  It includes all points where the orbital plane does not cross the line of sight, plus those connected orientations which never have a null in the waveform
amplitude.  Due to a coordinate degeneracy in Figure  \ref{fig:RegionDividers}, the region shown corresponding to R1
consists of two disjoint  mirror-image copies  R1$_\pm$,  related by  $\beta \rightarrow \pi-\beta$.  These two regions
include the special cases of spins aligned or antialigned with the
total angular momentum, viewed along or opposite to the the total angular momentum.
Conversely, the region R3 corresponds to binaries and viewing geometries so $L$ winds around \emph{both} null lines.  
The region R3 includes the special case of a \emph{precession disk} ($\beta=\pi/2$), viewed favorably edge on
($\theta_J=\pi/2,\psi_J=0$), plus those connected orientations free from amplitude zeros.   The two  regions R2$_{\pm}$
correspond to the ways $L$ can wind around one or the other null line.     At every polarization except $\psi_J=0$, the regions R1 and R3 are separated by R2.  For $\psi_J=0$, the
three regions overlap at the zeros  of %
\begin{eqnarray}
\label{eq:dcut}
\cos 2\theta_J + 2\cos 2\beta +1 = 2 [\sin^2 \theta + \cos 2\beta] \, .
\end{eqnarray}

\section{Waveforms in constant precession cone limit}
\label{sec:Waveforms}

In this section we review  how precession modulates the amplitude and phase of the binary's gravitational-wave signal.  
We express
precession-induced amplitude and phase modulations in terms of a complex  factor $z$.  We show the gravitational-wave
phase has both oscillating and secularly increasing contributions from precession.  The secular effect arises as a
fraction of the precession phase is imprinted on the signal.  We demonstrate that each of the
previously identified regions corresponds to a different amount of secular precession contribution.  To leading
order,  gravitational-wave detection with non-precessing templates works by matching a signal with similar \emph{phase}
evolution.  In this sense, gravitational-wave signals from each of the three regions are phenomenologically similar (and
can be phenomenologically distinguished from one another), as each region has a characteristic secular phase evolution
change from precession.

Explicit expressions for the  leading-order gravitational waveforms produced by instantaneously stationary orbits  are available in the
literature;  see, e.g.,  Eqs. (31-37) and Fig. 1 from \citet{2003PhRvD..67j4025B}.  
Additionally, though these expressions can be evaluated in any coordinate system (see, e.g., Fig. 1 in
\citet{2003PhRvD..67j4025B}), we adopt a specific coordinate system aligned with the total angular momentum.  
(These coordinates are different from the frame adopted in Eq. (\ref{eq:Coordinates:J}).)  
In our
coordinate system, the
expressions provided in \citet{2003PhRvD..67j4025B} correspond to the expressions below, with  their $\iota$ corresponding
to our $\beta$ and their $\Theta$ corresponding to our $\theta_J$.  In these expressions, the angle $\alpha$ is the
accumulated precession phase $\int \Omega_p dt$ of the orbital angular momentum around $\hat{J}$.   The reference  phase
$\Phi_S$  used in BCV is given by $\phi_{orb}-\alpha \cos\beta$ for $\phi_{orb}$ the orbital phase:  %
\begin{widetext}
\begin{subequations}
\label{eq:h:ViaCpSpEtc}
\begin{eqnarray}
h &\propto &(C_+F_+ + C_\times F_\times) \cos (2\phi_{orb}-2\alpha \cos \beta) 
+  (S_+F_+ + S_\times F_\times)\sin (2\phi_{orb}-2\alpha \cos \beta) \\
&\propto & \text{Re}[ (C_+F_\times + C_\times F_\times) - i (S_+F_++S_\times F_\times)]
\times  e^{2i(\phi_{orb}-\alpha \cos     \beta)} \\
C_+ &\equiv& \frac{1}{2} \cos^2 \theta_J [\sin^2 \alpha - \cos^2 \beta \cos^2\alpha] 
+ \frac{1}{2}(\cos^2 \beta \sin^2 \alpha - \cos^2 \alpha)
- \frac{1}{2}\sin^2\theta_J \sin^2 \beta - \frac{1}{4} \sin 2 \theta_J \sin 2 \beta \cos \alpha \\
S_+ &\equiv& \frac{1}{2}(1+ \cos^2\theta_J)\cos  \beta \sin 2\alpha + \frac{1}{2} \sin 2\theta_J \sin \beta \sin \alpha \\
C_\times &\equiv& - \frac{1}{2}\cos \theta (1+\cos^2 \beta) \sin 2\alpha - \frac{1}{2}\sin \theta_J \sin 2\beta \sin \alpha \\
S_\times &\equiv&  - \cos \theta_J \cos \beta \cos 2\alpha - \sin \theta_J \sin \beta \cos \alpha
\end{eqnarray}
\end{subequations}
\end{widetext}
with a corresponding expression for the orthogonal polarization. 
For the calculations needed here, however, the waveform is most usefully expressed as amplitude and phase modulation of a sinusoid: 
\begin{eqnarray}
h&\propto& \text{Re}[A e^{2 i\Phi_{wave}} ] = \text{Re}[z e^{2 i ( \phi_{orb} -\alpha \cos \beta)}] \\
\Phi_{wave}
&=& \phi_{orb} - \alpha \cos \beta + \frac{1}{2} \text{arg} z \\ %
\label{eq:def:zcomplex}
z&\equiv& (C_+F_++C_\times F_\times) - i (S_+F_+ + S_\times F_\times)
\end{eqnarray}
Practically speaking, the phase  increases and oscillates with time.   The gravitational-wave phase $\Phi_{wave}$ can be
decomposed into  three parts:
\begin{eqnarray}
\label{eq:PhaseDecomposition}
\Phi_{wave} &\equiv& \phi_{orb}  + W \alpha + \delta \phi_{prec} 
\end{eqnarray}
These three parts are  (a) orbital modulation $\phi_{orb}$; (b) a precession-induced secular increase in phase imposed by
geometric effects associated with precession ($W\alpha$); and (c) residual modulations in phase caused by precession
($\delta \phi_{prec}$).   
Both the secular evolution factor $W$ and modulations depend on the line of sight; see  Figure
\ref{fig:ThreeCanonicalPhasings} for three examples.

The secular phase evolution $W$ is particularly simple.  For example, for binaries and viewing orientations where the
orbital plane never crosses the line of sight, corresponding to  a single helicity being present along the line
of sight, the secular phase must accumulate steadily in proportion to this helicity.   Using the special case $\theta=0,\pi$ to determine suitable coefficients, one can show that
\begin{eqnarray}
W  &=& (\text{sign}(\cos\beta)-\cos\beta) 
\end{eqnarray}
More generally,  the secular phase factor $2W$ is $-2\cos \beta$ plus an \emph{integer}, related to the number of times the
complex number $z$
 winds around the point $z=0$ in a
precession cycle (i.e., versus $\alpha$):
\begin{eqnarray}
W &=& \frac{1}{2} n_{wind} - \cos \beta  = - \cos \beta + \frac{1}{4\pi i} \int \frac{d z}{z}
\end{eqnarray}   
where this integral follows a contour in $z$ corresponding to  $\alpha$ evolving  from $0$ to $2 \pi$ (i.e., one
precession cycle).

\begin{figure*}
\includegraphics[width=0.3\textwidth]{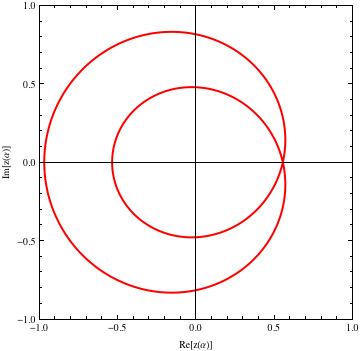}
\includegraphics[width=0.3\textwidth]{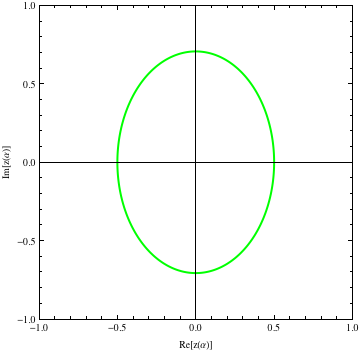}
\includegraphics[width=0.3\textwidth]{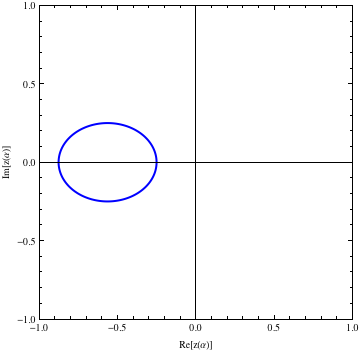}
\caption{\label{fig:z_alpha} \textbf{Complex amplitude contours}: Three examples of the complex amplitude $z(\alpha)$
  over a precession cycle, in the constant precession cone approximation.   The three configurations shown correspond
  to the three fiducial precession cones shown in Figure \ref{fig:ConeDemo}.
}
\end{figure*}

Figure \ref{fig:z_alpha} show examples of different trajectories for the complex
amplitude $z(t)$.   Depending on how the angular momentum winds around the two ``null lines'' [Figure \ref{fig:ConeDemo}], the complex number $z$
derived from the angular momentum and the detector orientation can wind around the origin (the point corresponding to
$\hat{L}$ along those null lines).
This integral can be explicitly evaluated, for example, by substituting $u=e^{i\alpha}$ and performing  a contour integral.
Since by construction the complex number $z$ is exactly zero at some point on its cycle if and only if
Eq. (\ref{eq:TransverseCondition}) holds,
the connected subsets of  R1, R2, and R3 are also regions of constant winding number $n_{wind}$.   Keeping in mind the
regions have different connected subsets depending on the sign of $\cos \beta$, we evaluate $W$ and find
\begin{eqnarray}
\label{eq:def:W:Discrete}
W&\equiv& -\cos \beta +\begin{cases}
\text{sign}(\cos \beta)  & \text{R1}  \\
\frac{1}{2}\text{sign}(\cos \beta) & \text{R2} \\
0 & \text{R3}
\end{cases}
\end{eqnarray}
\abbrvACST{}  find an approximately comparable  result [their Eq. (65); see also and \cite{1995PhRvD..52..605A}] by integrating an ODE for the phase perturbation due to
precession.  That expression does not include the region R2 or correctly identify the conditions that cause transition
from one secular phase to another. 
The complex amplitude $z(t)$ also encodes the way that the gravitational-wave phase and amplitude
``oscillate'' about their representative values. 
Binaries whose angular momenta are aligned with their orbit do not precess, so they do not accumulate secular precession
phase ($W=0$, as $z\propto e^{i2 \alpha}$) and their phase does not oscillate.  
 As is apparent from the $z(t)$ trajectories  in Figure
  \ref{fig:z_alpha}, any oscillations in amplitude and phase are more extreme for trajectories passing closer to
$z\simeq 0$.   These extreme oscillations cannot be well-fit by a non-precessing template.  We will show below that the
best fitting factor between a precessing signal and a non-precessing template family correlates directly with the amount
of phase and amplitude oscillation a signal exhibits.

 By way of example, Figure \ref{fig:ThreeCanonicalPhasings} shows how
precession introduces additional secular and oscillating contributions to the gravitational wave phase of the \emph{same}
binary, seen along different lines of sight.  
The three curves in this figure show $\text{arg}(z)$ versus $\alpha$, as extracted from $z(\alpha)$ [Figure
  \ref{fig:z_alpha}].
We can clearly see the distinct secular phase trajectories, corresponding to each of the three ways $\vec{L}$ can wind
around the ``null lines'' (equivalently, how $z(t)$ can wind around the origin).  

\section{Averaging amplitude and mismatch with the precession cone}
\label{sec:Averaging}

In this section we estimate the response of gravitational-wave detectors and non-precessing data analysis strategies to a
precessing signal.  Using a separation-of-timescales argument, we show that  precession-induced modulations
\emph{decouple} from the orbital phase, allowing key expressions like the signal amplitude and mismatch against proposed
signal templates to be computed by an average over the precession cone.  We apply this idea to compute how precession
modifies two quantities: the signal strength (SNR) and the mismatch of a non-precessing template with a precessing
signal.
Combined, these expressions tell us the relative signal strength a data analysis pipeline would recover from a
precessing binary, given we know how well it performs for a non-precessing counterpart (i.e., one with similar masses but aligned spins).

The signal-to-noise ratio for the detection of an exact template waveform
$h_e$ using an optimal filter constructed from a model waveform  $h_m$ used as
a template is given by
\begin{equation}
\rho_m = \langle h_e | \hat{h}_m \rangle = \frac{\langle h_e | h_m \rangle}{\langle h_m | h_m \rangle^{1/2}},
\end{equation}
where the noise-weighted inner product $\langle h_e | h_m \rangle$ is given by
\begin{equation}
\langle h_e | h_m \rangle = 2\int_{-\infty}^\infty df \frac{\tilde{h}_e(f)  \tilde{h}_m^\ast(f)}{S_n(|f|)}.
\end{equation}
Here $S_n(|f|)$ is the one-sided power spectral density of the detector strain
noise $n(t)$, defined by
\begin{equation}
\left<\tilde{n}(f) \tilde{n}^*(f')\right> = \frac{1}{2}\delta(f-f') S_n(|f|),
\end{equation}
and $\tilde{h}(f)$ is the frequency-domain representation of the waveform,
given by
\begin{equation}
\tilde{h}(f) = \int_{-\infty}^\infty h(t) e^{-2\pi i f t}\, dt \; .
\end{equation}

If we know the exact waveform, the strength of the signal is given by $\rho^2
= \langle h_e | h_e \rangle$. However, if there is an error in the waveform
model $h_m$, then the signal-to-noise ratio is reduced by a factor
\begin{equation}
\rho_m = M \rho
\end{equation}
where $0 \le M \le 1$ is the match of the model signal. The model waveform 
$h_m= h_m(\vec{\lambda})$ is a function of the model's intrinsic parameters
$\vec{\lambda}$ (e.g. masses, spins, etc.) and extrinsic parameters $\vec{\lambda}_e$ (e.g., time of arrival,
coalescence phase, sky location, etc.). The model waveform having the same
physical parameters as the exact waveform $h_m$ might not have the highest
match, and so we define the fitting factor $F$ as
\begin{equation}
F = \max_{\vec{\lambda},\vec{\lambda}_e} \langle\hat{h}_e | \hat{h}_m(\vec\lambda, \vec{\lambda}_e)\rangle.
\end{equation}
For the best-fit parameters, the signal-to-noise ratio will be reduced by a
factor of $\rho_m = F \rho$.

In this section we use the constant precession cone approximation to derive  closed-form expressions
for the fitting factor $F$ and the geometrical factors in the signal amplitude $\rho^2$.
Formally, our approximation arises through separation of timescales.
At each instant, the gravitational-wave emission is instantaneously quadrupolar.  
Expanding in powers of $\Omega_p/\pi f$, the time-domain amplitude modulation $A(t)$ implied by the beam pattern
 translates directly to
frequency-domain modulation $A(t(f))$ [\abbrvACST{} Eq. (36-37)]; see, e.g., Figure
\ref{fig:Simple:AmplitudeModulationIsEasy}.
Each frequency identifies an orientation of the angular momentum vector $\hat{L}(f)$ and thus precession phase $\alpha$.
So long as a few precession cycles dominate peak emission, a single (effectively ``constant'') precession cone dominates
as well.  Adopting this precession cone as fiducial, we can transform the inner products involving $A(t(f))= A_o(f)
B(\hat{L}(t))$ into inner products involving the ``slow'' degrees of freedom (i.e., just $A(f)$) and \emph{averages}
over the ``fast'' (precession) degrees of freedom ($B(\hat{L})$).   The averages over slow degrees of freedom do not
explicitly depend on precession geometry, detector noise, or component masses.  The expression factors.  We therefore arrive at expressions for $\rho^2$ and $F$ in
terms of averages.
As these averages encode all orientation-dependent effects, we emphasize the averages below, keeping in mind that their
prefactors (if needed) can be easily computed once and for all, using a face-on binary.

\begin{widetext}
\subsection{Averaged amplitude}

\begin{figure}
\includegraphics{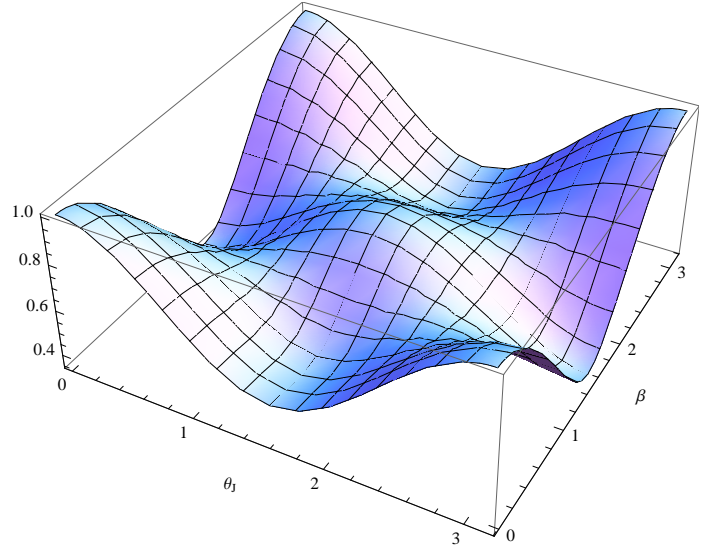}
\caption{\label{fig:AmplitudeModel:Geometric}\textbf{Amplitude ratio model}: For spinning precessing binaries, the
  function $s(\theta_J,\beta,\psi_J=0)$ describing the average amplitude along a given line of sight, assuming a
  \emph{steadily precessing} binary with precession cone opening angle $\beta$.  The function $s(\theta_J,\beta,\psi_J)$ [Eq. (\ref{eq:def:s})] is approximately the ratio
  $\rho/\rho_{ns}$ between the amplitude $\rho$  of a spinning, precessing binary seen in-band in a configuration and line of
  sight specified by $(\theta_J,\beta,\psi_J)$ and the amplitude $\rho_{ns}$ of a \emph{nonspinning} binary of comparable masses.  
}
\end{figure}

The precession of
the beam across the line of sight modulates the amplitude  seen at a single detector, in both $\sin$ and $\cos$ quadratures.  
We estimate the signal power by adding each quadrature's contribution independently, including the prefactor implied by
the orientation of $L$ relative to the line of sight and detector.  We then average the resulting expression over the
entire precession cone.  Compared to the geometrical factor for an optimally
oriented non-precessing source, the amplitude $\rho^2$ is smaller by a factor $s^2$:
\begin{eqnarray}
\label{eq:def:s}
s^2(\theta_J,\beta,\psi_J)&=&
  \left< \frac{(1+(\hat{L}\cdot n)^2)^2}{4}\cos^22\psi_L(t) + (\hat{L}\cdot n)^2 \sin^2  2\psi_L(t) \right> %
= 
 \left< \frac{(1+(\hat{L}\cdot n)^2)^2}{4} - \frac{(1-(\hat{L}\cdot n)^2)^2}{4} \sin^2  2\psi_L(t) \right> \nonumber \\
&=&  \left< \frac{(1+(\hat{L}\cdot n)^2)^2}{4}\right> - 
  \left< (\hat{L}\cdot \hat{x})^2(\hat{L}\cdot \hat{y})^2\right> \\
\label{eq:def:shat:exact}
&=& \frac{1}{1024}
\left[ \{c_p(x-1)^2 + x^2\}(35 y^2 + 10 y -13) + 2x \left(5 y^2+166 y+53\right)-13 y^2+106 y+451
 \right]
\end{eqnarray}
where $x=\cos 2\theta_J,y=\cos 2\beta$ and $c_p=\cos 4\psi_J$.
 Note that  all polarization dependence enters through a term $\propto c_p(1-x)^2 + x^2$; the average amplitude is
almost polarization-independent near $\theta_J \simeq 0$.

Figure \ref{fig:AmplitudeModel:Geometric} shows how the relative amplitude of a BH-NS binary seen directly overhead a
single detector should change, versus the purely geometrical orientation parameters $\theta,\beta,\psi_J$ describing the
constant precession cone.

\end{widetext}

\subsection{Averaging and fitting factor}

The phase evolution is more subtle.   As simple precession  parallel-transports the orbit along a precessing path, the gravitational-wave phase
$\phi_{wave}$ of the \emph{single polarization to which our detector is sensitive} oscillates about a combination of the orbital phase and a 
secularly accumulating  proportion of the precession phase $\alpha$:
\begin{eqnarray}
\label{eq:SimplePrecession:Phase:Secular}
2\phi_{wave} &=& 2\phi_{orb} + 2\delta\phi_{prec}(\alpha)   + 2  W \alpha
\end{eqnarray}
as described above.
Finally, the spin also influences the accumulation of orbital phase $\phi_{orb}$:  the
projection of $\vec{S}_1$ along $\vec{L}$ produces a ``gravitomagnetic repulsion'', changing the orbital evolution
$\phi_{orb}$ by slowing the rate of increase in frequency; see, e.g., Eq. (6) in  \cite{BCV:PTF}.    

For  detectors like LIGO, which are sensitive to an epoch where the binary's opening angle is nearly constant, 
the \emph{secular} phase changes ($\phi_{orb}$ and $W \alpha $) can be well-fit by  a \emph{sufficiently generic}
non-precessing waveform.\footnote{  Roughly speaking this ansatz requires  a large enough binary model space and small enough time range that the equation
    $\phi_{orb}(\lambda,f)+W(\lambda)\alpha(f) \simeq \phi_{orb}(\lambda',f)$ can be solved for one single set of
    parameters $\lambda'$, for all $f$ in band and for each $\lambda$.  Empirically, this ansatz works well, even when the non-precessing templates have no spin.   To
    be concrete, we performed Monte  Carlo studies described in Figure \ref{fig:ProofItWorks:Generic} using a
  \emph{nonspinning} template bank.  So long as the template bank allowed the mass ratio parameter $\eta$ to be greater than 1/4, all aligned
  spins could be well-fit in the mass and mass ratio range of interest.    Moreover, by explicitly constructing a
  waveform without geometrically-induced modulations, we have confirmed a good fit between the secular evolution
and non-precessing signals.}      For example, the secular phase evolution of many precessing waveforms can be well-fit by
a non-precessing waveform with different chirp mass and mass ratio ($\mc,\eta$).   On the other hand, amplitude and phase oscillations in $A(t)$ and $\delta \phi_{prec}$
cannot be fit with non-precessing templates.  

  Using the ansatz that the
best non-precessing template $h_m$ will perfectly reproduce the \emph{secular} phase,  the
time-averaged overlap between the spinning waveform $h_e$ and the best nonspinning template must be proportional to the best nonspinning
fitting factor.  We therefore define approximation to the fitting factor $F$ that
captures the effects of precession by
\begin{eqnarray}
s \mathcal{F} \equiv \int \frac{dt}{T}\qmstateproduct{h_e(t)}{h_m(t)} 
\end{eqnarray}
where $s$ is the average amplitude from Eq. (\ref{eq:def:s}) and
where the average is over a single precession cycle of $\hat{L}$ in $h_e$, and where $h_m$ is the best-fitting  nonspinning
template to the entire cycle.   
Replacing a time average by an average over the precession phase $\alpha$ and the true signal $h_e(t)$ with
the nonspinning template times the amplitude and (fourier domain) precession phase terms  $A$ and $\delta \phi$ described
above, this expression reduces to an average of the oscillating amplitude and phase over the cycle:
\begin{eqnarray}
s \mathcal{F} &=& \text{max}_{\phi_0} \int A \cos (2\delta \phi+\phi_o) d\alpha/(2\pi)
\end{eqnarray}
We will provide a closed-form expression in the next section.  
However, to  understand that result, it is helpful to think directly in terms of this integral.  
For example, in the limit of small amplitude and phase modulations occurring periodically on the precession cone, a
non-precessing template model should fit them to no better than 
\begin{eqnarray}
\mathcal{F}&\simeq&
  1 
 -\frac{1}{2}
   \left[
    \left<(2\delta \phi)^2\right> + \left<(\delta A/A)^2\right> - \left<\delta A/A\right>^2
   \right] 
\end{eqnarray}
[This expression can be rederived directly from the original integral.]     
In particular, the amplitude and phase modulations introduced by precession inevitably diminish the ability of a
non-precessing template to match them, by an amount proportional to the standard deviation of oscillations in time and
phase about the reference model.

Oscillations in phase and amplitude are most extreme for binary geometries near the boundaries between regions. 
A non-precessing signal should be maximally unable to match the large phase and amplitude variations near this surface.  
For example, a signal on the surface has different epochs, separated by zeros of the (complex) amplitude.   A
non-precessing signal can coherently reproduce the  phase trend in one portion of the signal or the other, but not both. 

\subsection{Constant Precession Cone}

In the limit of a constant precession cone, the integral can be performed analytically.  
Replacing a time average by an average over the precession phase $\alpha$ and the true signal $h(\hat{L})$ with
the nonspinning template times the amplitude and (fourier domain) precession phase terms  $A$ and $\delta \phi$ described
above, the angle-averaged expression reduces 
\begin{eqnarray}
s \mathcal{F} &=& 
  \sqrt{(s\mathcal{F}_s)^2 + (s\mathcal{F}_c)^2} 
\end{eqnarray}
where in the second line we convert maximization of the integrand over $\phi_o$ into two integrals.
\begin{eqnarray}
s\mathcal{F}_c &\equiv & \int A_{tot} \cos 2\delta\phi \frac{d\alpha}{2\pi} \\
s\mathcal{F}_s &\equiv & \int A_{tot} \sin  2\delta \phi  \frac{d\alpha}{2\pi} 
\end{eqnarray}
This expression reduces to a simple contour integral: 
\begin{eqnarray}
s\mathcal{F}_c &\equiv & \int \frac{(z/Z+(z/Z)^*)}{2}\frac{d\alpha}{2\pi}   \\
s\mathcal{F}_s &\equiv &\int \frac{((z/Z)-(z/Z)^*)}{2i}\frac{d\alpha}{2\pi}   \\
z(\alpha)&\equiv & A_{tot}e^{i2(\delta \Psi + \alpha W) }  \\
Z(\alpha)&\equiv& e^{i 2 W \alpha }
\end{eqnarray}
where $z$ is the complex representation of the waveform amplitude and phase given in Eq. (\ref{eq:def:zcomplex}), while
dividing by $Z$ is equivalent to subtracting
the secular phase.  Using the complex
variable $u=e^{i\alpha}$ to represent $Z=u^{2W}=u^{n_{wind}}$ and  the sinusoids in $C_{+,\times}$ and $S_{+,\times}$ analytically  via
$\cos \alpha = (u+1/u)/2$ and similarly,
 we find the fitting factor predicted  can be calculated using the contour integral
\begin{eqnarray}
s\mathcal{F} &=& |I| \\
I &\equiv& \int\frac{z/Z d\alpha}{2\pi} = \int \frac{z/u^{n_{wind}} du}{2\pi i u} 
\end{eqnarray}
This trivial contour integral corresponds to identifying terms  $\propto u^n$ in $z$ for different $n$.   
The value of this contour integral depends discontinuously on the winding number $n_{wind}$.  Using the definitions of
$C,S$ to perform this integral, we find  
\begin{widetext}\begin{subequations}
\label{eq:def:MatchEstimate:ClosedFormViaContour}
\begin{eqnarray}
I &\equiv& \begin{cases}
-\frac{3}{4} \cos 2 \psi_J \sin^2 \beta \sin^2 \theta & n_{wind}=0 \\
\mp \frac{(2\sin\beta \pm \sin 2\beta)( \cos 2\psi_J \sin 2\theta \mp  2i \sin \theta \sin 2\psi_J)}{8} & n_{wind}=\pm 1 \\
-\frac{(1\pm \cos \beta)^2}{8} \left [ \cos 2\psi_J (1+\cos^2\theta)  \mp  2 i \cos \theta \sin 2\psi_J \right] & n_{wind}=\pm 2
\end{cases}
\end{eqnarray}
Finally, we estimate the fitting factor for all angles by dividing this closed-form expression by the closed-form expression
for $s$:
\begin{eqnarray}
\mathcal{F}(\theta_J,\beta,\psi_J)&\equiv& |I(\theta_J,\beta,\psi_J)|/s(\theta_J,\beta,\psi_J)
\end{eqnarray}

\end{subequations}
\end{widetext}

This ratio varies strongly in all three parameters and is best summarized by the explicit expression above.
Though complicated, some important physical lessons can be drawn.  First and foremost, for non-precessing signals the predicted fitting factor is $F=1$.    Second and most critically, while the
predicted fitting factor has a local extremum with $\mathcal{F}\simeq 1$ in each of the three regions, the predicted fitting factor drops rapidly as we
approach the surface dividing them: at the surface defined by Eq. (\ref{eq:def:dnull:zeros})  and shown in Figure \ref{fig:RegionDividers}.  In other words, this expression suggests the worst possible matches will occur when
the angular momentum at some point  in the precession cycle is along one of the two null lines. 
As noted above, the low fitting factor in this region occurs because the signal consists of comparable amounts of power with two
competing and incompatible secular phase trends.  
Finally, under special conditions such as $\theta\simeq 0$ in R2 and R3, this expression predicts pathologically low
matches $P\simeq 0$.  As we will discuss at greater length in a forthcoming publication, an empirically more successful
approximation replaces the options in $I$ by the the maximum over the three options.

\begin{figure}
\includegraphics[width=\columnwidth]{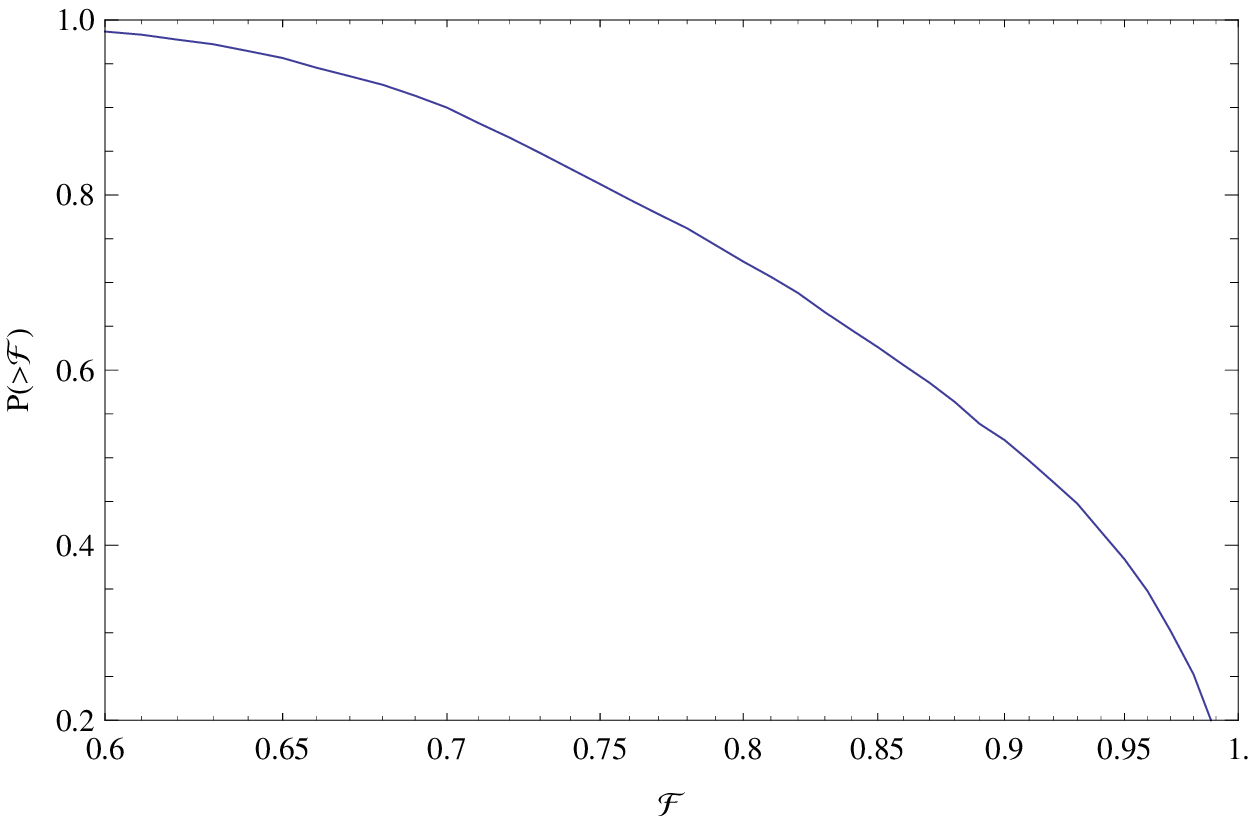}
\caption{\label{fig:SampleSyntheticMonteCarlo} \textbf{Distribution of fitting factors for single-spin BH-NS}: Using
  only the analytic expressions provided in this paper, the predicted fraction $P(>{\cal F})$ of fitting factors greater
  than ${\cal F}$ for a   randomly oriented BH-NS binary with $m_1=10 M_\odot,m_2=1.4 M_\odot$, with the dimensionless black hole spin $\vec{\chi}_1$ randomly oriented and with
  magnitude uniformly distributed between $0$ and $0.98$.   Compare to the solid line in the bottom right panel of Figure 9 in \cite{2011PhRvD..84h4037A}, although there the neutron star also has a spin up to $0.3$.
}
  \end{figure}

Though derived under the seemingly-strong expression of a constant precession cone, we have verified this
expression applies to a broad family of real BH-NS inspirals,  by calculating the overlap between the gravitational radiation emitted from
non-precessing and precessing binaries and maximizing over the non-precessing signal's parameters.  
We will describe these Monte Carlo studies in more detail in a subsequent publication.   In brief, these studies
corroborate many earlier studies that examined how well non-precessing signals recover precessing waveforms
\cite{1995PhRvD..52..605A,2003PhRvD..67j4025B,2011PhRvD..84h4037A}.   First, in special but not uncommon circumstances,
fitting factors  as low as $0.6$ occur.  Second, small fitting factors (i.e., large mismatches) can occur even with
small amounts of spin-orbit misalignment, though they become more common as misalignment increases.   For nearly aligned
binaries, signals with the lowest match correspond to directions with low signal power.  Contrary to previous claims,
however, we identify some configurations with both low fitting factor and significant signal power along the line of
sight.  
This analytic  study differs from all previous investigations in that we assume a non-precessing signal model can always
match any secular phase evolution; identify which precessing configurations lead to large
 modulations; and quantify the expected purely geometrical effect these modulations have on the fitting factor.
As an example, in Figure \ref{fig:SampleSyntheticMonteCarlo} we predict the distribution of fitting factors expected
when a non-precessing search is applied to randomly-oriented BH-NS binary.   Despite the simplicity of our model, we
recover quantitatively similar results to previous studies \cite{2011PhRvD..84h4037A}.  
Unlike earlier studies, however, our expressions allow us both to identify precisely which binaries are least well fit
and to understand precisely why.

\subsection{Validating our approximations with Monte Carlo}

To demonstrate that our approximation works, we have compared randomly-chosen BH-NS binaries with a standard non-precessing
template bank: Taylor F2 templates with physical parameters, hexagonal template bank chosen for 97\%
  minimal match; and a likely early advanced LIGO noise curve, the low-power zero-detune PSD \cite{LIGO-aLIGODesign-Sensitivity} such that $f_{peak}\simeq 40$ Hz.    

\begin{figure}
\includegraphics[width=\columnwidth]{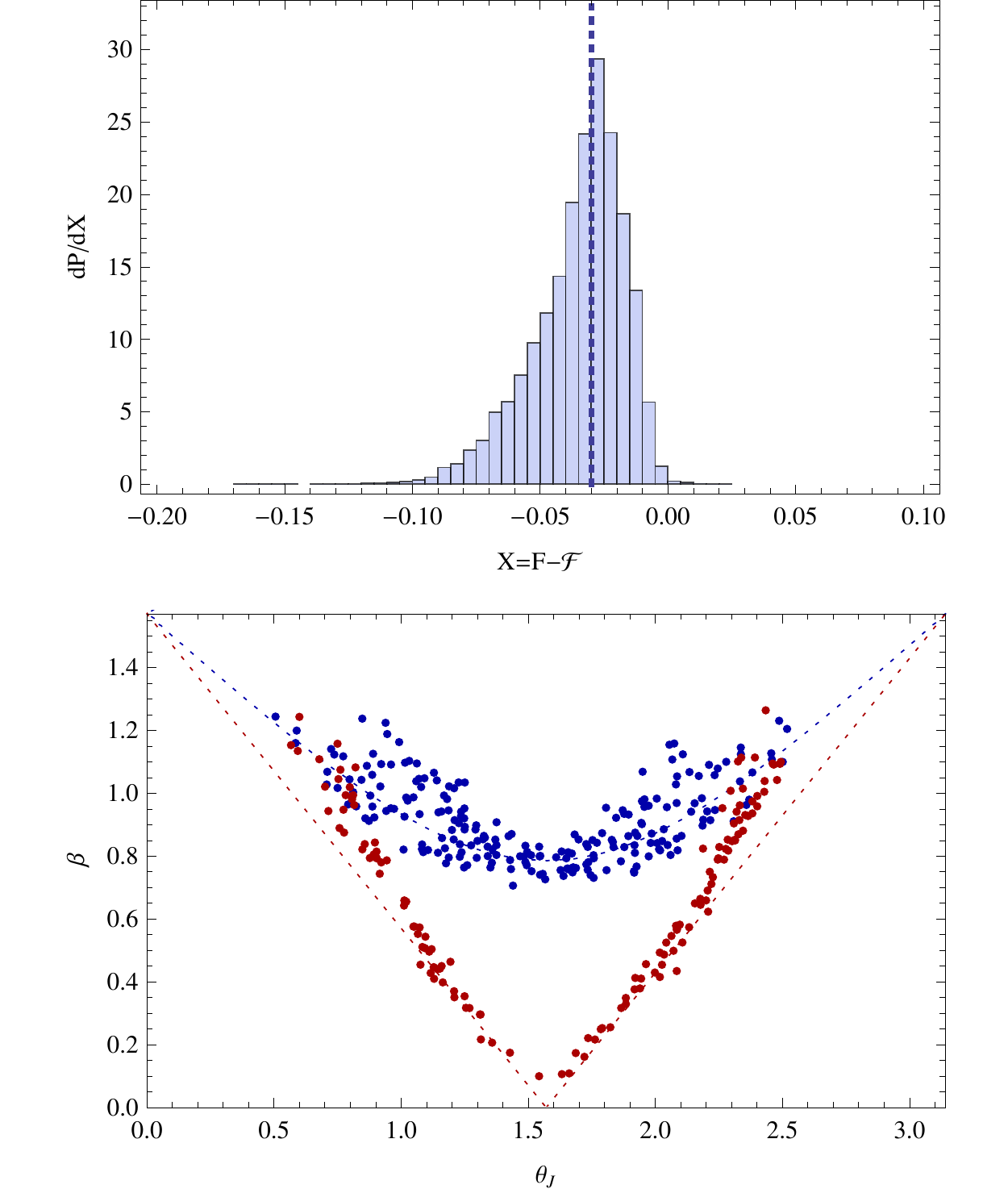}
\caption{\label{fig:ProofItWorks:Fiducial}\textbf{Comparison of the analytic prediction of Sec. V.C with numerical results I}: Single binary
  masses and spin magnitude \emph{Top panel}:
  For 30000 BH-NS binaries ($10M_\odot+1.4M_\odot$ with $a_1\in[0,1]$) with random orientations for $\hat{L}\cdot
  \hat{S}>-1/2$, a comparison between our estimate ${\cal F}$ and the fitting factor $F$ calculated by comparing that signal with a standard
  single-detector template family (Taylor F2 templates with physical parameters;  hexagonal template bank chosen for 97\%
  minimal match; early-stage advanced LIGO noise curve  such that $f_{peak}\simeq 40$ Hz \cite{LIGO-aLIGODesign-Sensitivity}).   For comparison, the vertical dotted line at $-0.03$ shows the
  typical mismatch expected from template bank discreteness.    
Despite known systematic differences between the signal and template model even for aligned systems,
  our match prediction works well almost everywhere.    To highlight its effectiveness, we have eliminated the
small fraction of  worst-performing configurations, corresponding to sources recovered with templates  with  $0.24 < \eta \leq 1/4$.  Due to the finite parameter extent of physical Taylor F2 templates, the conventional signal model cannot
  reproduce large aligned spins.     
\emph{Bottom panel}: For the same comparison of template bank against signals  as above, the configurations of the worst-fit sources ($F<0.7$) with $\cos 4\psi >
0.9$ (blue; $J$ nearly aligned with the detector on the plane of the sky) and $\cos 4\psi < -0.9$ (red; $J$ nearly
$45^o$ off the detector arms, in the plane of the sky).   For comparison, the blue and red solid lines are the surfaces
$\cos 2 \beta+ \cos^2 \theta=0$ (blue) and $\cos 2\theta_J + 2\cos 2\beta$ (red) [Eq. (\ref{eq:dcut})].   These lines
are two one-dimensional cuts through the surface separating  R1 and R2, where our expression predicts the worst
single-detector matches will occur.}
\end{figure}

For example, Figure \ref{fig:ProofItWorks:Fiducial} shows the results for  30000 BH-NS binaries ($10M_\odot+1.4M_\odot$) with random orientations for $\kappa = \hat{L}\cdot
  \hat{S} >- 1/2$ and spin magnitudes $a_1\in[0,1]$. The limit on $\kappa$ removes some systems with transitional precession for clarity; in the next figure, we allow nearly the full range of $\kappa$.
The bottom panel shows the configurations of the worst-fit sources for two polarizations ($J$ nearly aligned with the
detector on the plane of the sky and $45^o$ off the detector arms).    Both cluster near the predictions of our model
(dotted lines): our approximation correctly identifies the worst-fit locations.
More generally, the top panel shows our estimate works (nearly) everywhere: it compares  our fitting factor estimate
${\cal F}$ to  the  fitting factor $F$
  calculated by comparing that signal the template bank.   Despite known systematic differences between the signal and template model even for aligned systems,
  our match prediction works well almost everywhere.     For comparison, the vertical dotted line at $-0.03$ shows the
  typical mismatch expected from template bank discreteness.    

\begin{figure}
\includegraphics[width=\columnwidth]{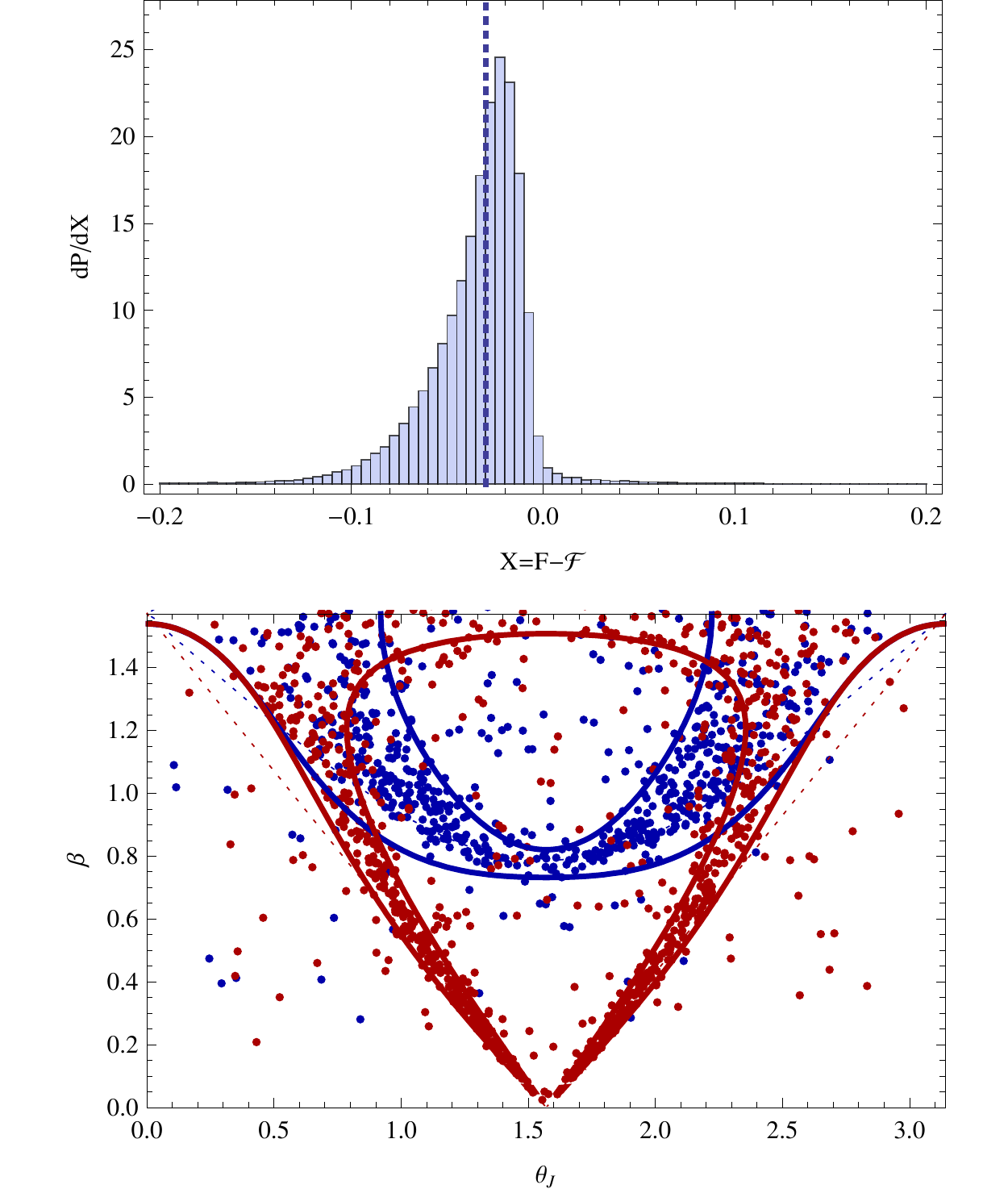}
\caption{\label{fig:ProofItWorks:Generic}\textbf{Comparison of the analytic prediction of Sec. V.C with numerical results II}: Monte Carlo.  As Figure \ref{fig:ProofItWorks:Fiducial}, except (a) the injected signal can have
  generic BH-NS-like masses $m_1\in[3,15] M_\odot$, $m_2\in[1,5] M_\odot$, dimensionless spin magnitudes $\chi \in[0,1]$, and spin-orbit
  misalignments $\hat{L}\cdot \hat{S} > -0.9$; (b) to better recover aligned-spin signals, the template family includes
  Taylor F2 waveforms with $\eta>1/4$, spaced in chirp mass and $\eta$ according to the 2PN hexagonal template bank. 
\emph{Top panel}: Our match prediction works for generic sources.  [Because the mass and spin distribution used in
  this analysis favors low $\gamma$ and thus binaries nearly aligned in band, this distribution partially reflects our ability to recover
  nearly-aligned-spin binaries with an extended $\eta$ Taylor F2 bank.  That said, subsets of the sample bounded below
  in $\gamma$ also demonstrate our prediction works well for generic binaries with significant in-band misalignment.]
\emph{Bottom panel}: Our expression identifies the dominant physical mechanisms that produce a poor fit between a
spinning binary and a candidate non-precessing waveform.
Like  Figure \ref{fig:ProofItWorks:Fiducial}, the dotted lines show the seperatricies between regions.  The \emph{solid
  lines} show the contours predicted from our expression for ${\cal F}=0.75$ and $\psi=0$ (blue) or $\psi=\pi/4$ (red).
For comparison, the points indicate all signal versus bank fitting factors $F<0.66$ for $\cos 4\psi >0.975$ (blue) or $F<0.7$ for $\cos 4\psi <-0.975$.
}
\end{figure}

In this comparison, we have artificially eliminated the
small fraction of  worst-performing configurations, corresponding to sources recovered with templates near the edge of the template bank where $\eta = 1/4$. The secular phasing due to the aligned component of the spin is best fit by templates with $\eta$ biased to larger than its true value. Due to the finite parameter extent in $\eta$ of physical Taylor F2 templates, the conventional signal model fails to capture even the secular effects of large aligned spins. An aligned but non-precessing template bank would however be able to caputre these signals.    
To faithfully reproduce signals with aligned spins using standard template banks, we must extend the bank into
$\eta>1/4$. This nearly captures the secular effect of aligned spins, and is a reasonably good subsitute for a true aligned spin template bank.  Figure \ref{fig:ProofItWorks:Generic} shows the results of this  extended-$\eta$ bank, compared to a much larger set of
BH-NS binaries. The top panel shows that our approximation continues to work well for generic sources, even given systematic differences
between signal and template and the absence of spin.    In particular, the bottom panel demonstrates that except for
rare outliers, our approximation correctly identifies the worst-fit sources.

\subsection{Beyond the constant precession cone}
\begin{figure}
\includegraphics[width=\columnwidth]{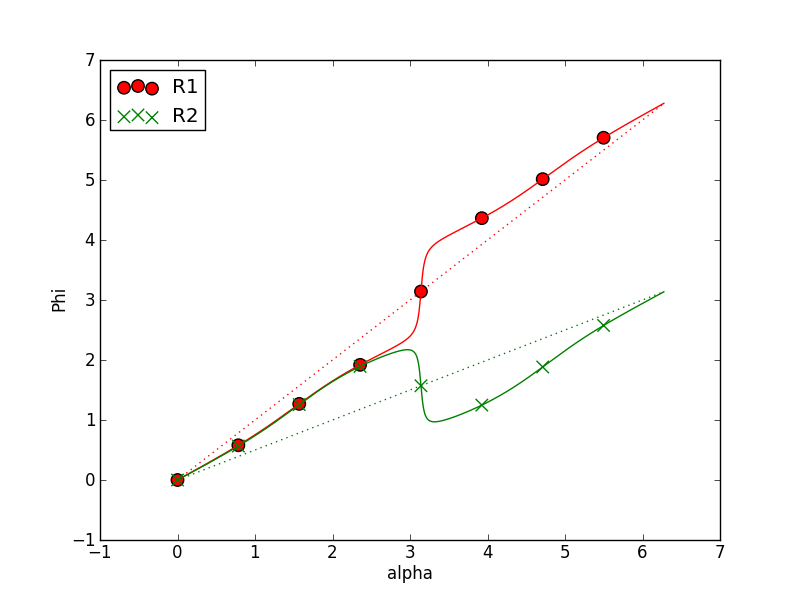}
\caption{\label{fig:PhasingTransitionFaceCross} \textbf{Discrete phase change}: With a fixed viewing geometry, the
  geometrical phase difference $\text{arg}(z)$ on either side of a transition surface (here,
  R1 to R2). As in Figure \ref{fig:ThreeCanonicalPhasings}, an entire precession cycle is shown.  }
  \end{figure}

Though we emphasize the value of a (nearly-)constant precession cone in approximating binary inspiral, the concepts and expressions we provide apply
equally well to many simply-precessing binaries.   As an example, in Figure
\ref{fig:PhasingTransitionFaceCross} we show the precession-induced phase evolution for two systems with slightly different $\beta$, such that they are near a transition between regions. One is in R1 and the other in R2. The nearness to the transition surface causes large oscillations around the secular increase in phase.   
 Binaries whose opening angles $\beta(f)$ evolve across these boundaries and particularly across $\pi/2$ in band can
 undergo dramatic transitions in the properties of their observable waveform,  accumulating only one helicity early on
 and another helicity later. 

\section{Conclusions}

In this paper we have explored two features of the gravitational-wave signal seen by ground-based detectors in a ``constant precession
cone'' limit.  We compare their amplitude to  similar face-on non-precessing binary; we compare their waveform to the
best-fitting non-precessing signal.   We find closed-form expressions for both quantities.  Our expressions involve
only geometrical parameters -- the orientation of the precession cone relative to the line of sight --  with no explicit reference to the (post-Newtonian) phase model,
detector noise curve, or binary parameters.
In a subsequent paper we will use these factors to interpret the selection biases of optimal and non-precessing searches
for precessing binaries.    This publication will provide explicit Monte Carlo over all possible source parameters and
sky locations and calculate overlaps using state-of-the-art signal models.

In this paper we provide only the leading-order geometric influence that the precession cone has on signal amplitude.
Additionally, the same spin misalignments that allow the binary's orbit to precess also diminish the binary's inspiral
time, compared to an aligned non-precessing signal.  We will describe a more complete model for the spin-dependent signal
amplitude $\rho^2$ in a future paper.

Finally, in this paper we point out that non-precessing signals often fit precessing signals' \emph{secular} phase evolution.
The secular evolution of (single-detector) gravitational-wave phase depends on the line of sight to the source and
orientation of the detector ($W$).
As a result, two independent detectors running two independent searches may identify two very different best-fitting
(non-precessing) signals, due only to the detectors' orientations.  These biases in recovered parameters must be taken
into careful account when constructing coincidence-based search algorithms. We will also address biases in recovered
parameters in a future paper.

\acknowledgments
The authors benefitted from helpful feedback from Tom Dent, Jacob Slutsky, and Evan Ochsner.
DB and AL are grateful to NSF award PHY-0847611 for support. Additionally, AL is grateful to NSF grant PHY-0855589. DB is supported
by a Cottrell Scholar award from the Research Corporation for Science
Advancement.  ROS is supported by NSF award PHY-0970074, the Bradley Program
Fellowship, and the UWM Research Growth Initiative. ROS was also supported by
NSF award PHY-0653462 and the Penn State Center for Gravitational Wave
Physics. Computations used in this work were performed on the Syracuse
University Gravitation and Relativity cluster, which is supported by NSF
awards PHY-1040231, PHY-0600953 and PHY-1104371.

\bibliography{paperexport}%

\end{document}